\documentclass[twocolumn]{aastex63}
\usepackage{natbib}
\usepackage{color}
\usepackage{mathtools}
\usepackage{booktabs}
\usepackage{hyperref} 
\usepackage{mathtools}
\usepackage{comment}

\DeclarePairedDelimiterX\braket[2]{\langle}{\rangle}{#1\,\delimsize\vert\,\mathopen{}#2}

\def    \apjl  		{\rm {ApJL}}

\def    \apjl  		{\rm {ApJL}}

\def	\cm		{\,{\rm {cm}}}
\def	\K		{\,{\rm K}}
\def	\g		{\,{\rm {g}}}
\def	\mum	{\,{\mu \rm{m}}}

\def \bea {\begin{eqnarray}}
\def \ena {\end{eqnarray}}

\def    \bB     {\bf  B}

\def	\bB	{\boldsymbol{B}}

\def	\bS	{\boldsymbol{S}} 
\def    \bmu    {{\hbox{\boldsym\char'026}}}	

\def	\bv	{\boldsymbol{\rm v}}

\def	\cm	{\,{\rm cm}}

\def	\max	{\,{\rm max}}
\def	\d	{{\rm d}}

\def	\D	{{\rm D}}

\def	\erg	{\,{\rm erg}}
\def	\eV	{\,{\rm eV}\,}
\def	\g	{\,{\rm g}}

\def	\G	{{\rm G}}

\def	\H	{{\rm H}}

\def	\N	{{\rm N}}

\def	\s	{\,{\rm s}}

\def	\sr	{\,{\rm sr}}

\def	\AU	{\,{\rm au}}

\def	\V	{{\rm V}}
\def	\Bar	{{\rm Bar}}

\def	\yr	{{\rm yr}}
\def	\IP	{{\rm IP}}

\def	\bS			{\boldsymbol{S}}

\def    \ev     	{{\rm eV}}

\def    \Im     	{{\rm Im}}

\font\mib=cmmib10

\def\bOmega{\hbox{\mib\char"0A}}
\def\bmu{\hbox{\mib\char"16}}

\makeatletter
\newcommand*{\rom}[1]{\expandafter\@slowromancap\romannumeral #1@}

\begin{document}
\shorttitle{Role of CRs on secondary spin-polarized electrons}
\shortauthors{Thiem Hoang}

\title{Spin-Polarized Electrons from Magnetically Aligned Grains and Chiral Symmetry Breaking: Effects of Cosmic Rays in Protostellar Environments}

\author{Thiem Hoang}
\affiliation{Korea Astronomy and Space Science Institute, Daejeon 34055, Republic of Korea} 
\email{thiemhoang@kasi.re.kr}
\affiliation{Department of Astronomy and Space Science, University of Science and Technology, 217 Gajeong-ro, Yuseong-gu, Daejeon, 34113, Republic of Korea}

\begin{abstract}
Low-energy spin-polarized electrons (SPEs) are thought to cause symmetry breaking and could explain the origin of homochirality of prebiotic molecules such as amino acids and sugars. Here we study the effect of cosmic rays (CRs) on the emission of SPEs from aligned grains in dense protostellar environments and explore their effects on chiral asymmetry of prebiotic molecules. We first show that icy grains in protostellar environments can align with magnetic fields due to magnetically enhanced radiative torque mechanism. We then study the production of thermal electrons by CR ionization of H$_2$ and the CR-induced UV radiation using the attenuated CR spectra in dense cores obtained from a continuous slowing down model. Next, we show that thermal electrons with initial random spins captured by aligned grains will become spin-polarized due to the Barnett effect, converting unpolarized electrons into SPEs. We calculate the rate of photoemission of such SPEs by CRs-induced UV radiation and secondary electron emission from aligned grains and find that the photoemission by CRs-induced UV radiation is dominant. Finally, we calculate the total production rate of SPEs inside aligned dust grains by CRs. We estimate the alignment degree of SPEs from superparmagnetic (SPM) grains and find that it is only significant for SPM grains having large iron clusters and fast rotation. We suggest that low-energy secondary SPEs from aligned superparamagnetic grains with large iron inclusions induced by CRs might cause the chiral asymmetry of chiral prebiotic molecules formed in the ice mantle of aligned grains, in analogous to UV circularly polarized light. We propose that if amino acids and sugars of chiral assymmetry detected in meteorite/asteroids/comets might be formed in icy grain mantles with superparamagnetic inclusions under the irradiation of SPEs released from aligned grains by CRs in protostellar environments.

\end{abstract}


\section{Introduction}
Homochirality of biological molecules (e.g., amino acids and sugars) is the unique biosignature of life on Earth, as first discovered in 1848 by Louis Pasteur (e.g., \citealt{Bonner.1991}). A chiral molecule has two non-superimposable mirror images, called enantiomers. Chemical reactions typically produce a racemic mixture of equal amounts of left-handed and right-handed enantiomers (e.g., Miller-Urey experiment, \citealt{MillerUrey.1959}). However, chiral molecules of the biological origin are composed of only one enantiomer (also called an optical isomer). For example, amino acids and proteins have only left-handed enantiomers, whereas sugars, DNA, and RNA have only right-handed enantiomers (see \citealt{Sparks.2015} for a review). \cite{Soai.1995} and \cite{Blackmond.2004} showed that an initial small enantiomer excess could be amplified by autocatalytic reaction and eventually for pure enantiomer.


Amino acids and sugars were also discovered in carbonaceous meteorites, including the Murchison meteorite \citep{Kvenvolden.1970,Cronin.1997,Pizzarello.1998}, the Murray meteorite \citep{Pizzarello.1998}, the Tagish Lake meteorite \citep{Glavin.2012}. Recently, some amino acids are discovered in the cometary coma of 67P/Churyumov-Gerasimenko in the Rosetta mission \citep{Altwegg.2016}. Most of those prebiotic amino acids from meteorite exhibit the L-enantiomer, the same as amino acids in biological molecules on Earth. Isotopic analysis of deuterium in meteorite \citep{Engel.1997,Pizzarello.2005,Glavin.2020,Glavin.2020wwc} supported the extraterrestrial origin of amino acids in carbonaceous meteorites. The remaining questions are whether amino acids can form in the interstellar medium (ISM) where new stars and planets are born, see \citealt{Glavin.2020} for a review), and how chiral symmetry of chiral molecules could be broken in the star- and planet-forming regions.


Complex organic molecules (COMs, e.g., C$H_{3}$OH and C$_{2}$H$_{5}$OH), the building blocks of amino acids and sugars, are believed to form in the ice mantle of interstellar dust grains from simple molecules such as H$_{2}0$, CO, HCN, and NH$_{3}$ \citep{Herbst:2009go,Caselli:2012fq}. Independent experiments by \cite{Bernstein.2002} and \cite{Caro.2002} demonstrated that amino acids (including glycine, alanine, and serine) could be formed from irradiation of UV photons on the analogs of interstellar ice (consisting of water (H$_2$O), hydrogen cyanide (HCN), ammonia (NH$_{3}$), methanol (CH$_{3}$OH)) (see also \citealt{Elsila.2007}). UV irradiation on interstellar ice analogs can also form sugars \cite{Meinert.2016}. The recent experiment in \cite{Ioppolo.2021} showed that glycine could be formed in the ice mantle without the need for energetic irradiation of UV radiation and CRs. \cite{Parker.2023} suggested the concentration of amino acids in the Ryugu asteroid could arise from the formation of amino acids in the interstellar icy grains or ice mantle of grains in the presolar nebula, which then is accreted onto planetesimals during planet formation. Therefore, one plausible route is that amino acids are formed in the ice mantle of interstellar grains via UV photochemistry rather than formation in liquid water on an early Solar System body \citep{Elsila.2007,Oba.2016,Modica.2018,Oba.2023,Potiszil.2023}. 

In astrophysical environments, the first interstellar amino acid (propylene oxide) was discovered toward the Galactic center by \cite{McGuire.2016} using the rotational spectroscopy observed with single-dish radio Green Bank Telescope (GBT). Recently, ethanolamine (a precursor of phospholipids, \citealt{Rivilla.2022}) and glycine isomer \citep{Rivilla.2023} were detected toward the Galactic Center using the rotational spectroscopy using Yerkes 40 m and IRAM 30 m radio telescopes. These observations demonstrated that prebiotic chiral molecules could be formed in astrophysical environments. 


UV circularly polarized light (UVCPL) is widely known to induce chiral assymmetry (i.e., enantiomer excess) due to selective destruction of one enantiomer of chiral molecules \citep{Flores.1977,Bonner.1991}. An experiment by \cite{Modica.2014} showed that amino acids formed in ice mantles by irradiation of UVCPL on interstellar ice analogs have enantiomer excess. Circular polarization of near-infrared light is observed in star-forming regions \citep{Bailey.1998,Kwon.2016,Kwon.2018}. Therefore, the differential absorption of UVCPL by chiral molecules is a plausible mechanism producing the initial enantiomer excess and enantio-enrichment in the ISM \citep{Bailey.2001}. The limitation of this mechanism is that UVCPL cannot penetrate deep into dense clouds where new stars and planets are forming disks, and the enantiomer excess by UVCPL can be destroyed by other dynamical processes. To explain the chiral asymmetry of amino acids in meteorites/comets/asteroids, the enantiomer excess by UVCPL must be reserved during the planet formation process.

Very recently, a new promising channel to form COMs is introduced, through irradiation of low-energy electrons instead of UV photons. Numerous experiments \cite{Boyer.2016,Sullivan.2016,Kipfer.2024} have demonstrated that irradiation of interstellar ice analogs with of low-energy electrons (energy below $\sim 10-20$ eV) could trigger chemical reactions in similar ways as irradiation of UV photons. In particular, the experiment by \cite{Esmaili.2018} showed that glycine can be formed by irradiation of low-energy electrons on interstellar analog of CO$_{2}$-CH$_{4}$-NH$_{3}$ ice (see \cite{Arumainayagam.2019} for a recent review on the role of photochemistry vs. radiation chemistry). 

On the other hand, recent experiments established the key role of low-energy ($\lesssim 10$ eV) spin-polarized electrons (SPEs) and ferromagnetic surfaces in producing chiral asymmetry (see reviews by \citealt{Naaman.2018,Rosenberg.2019}).\footnote{Early studies suggested that the chiral asymmetry might arise through the preferential destruction of one enantiomer in a racemic mixture by SPEs produced in the $\beta$ decay of radioactive nuclei in weak interactions \citep{LeeYang.1956,Wu.1957}. However, the induced chirality asymmetry was found to be rather weak \citep{Hegstrom.1980,Bonner.1991}).} For example, \cite{Rosenberg.2008} first demonstrated experimentally that low-energy SPEs resulting from irradiation of UV radiation on a magnetic substrate \citep{Kisker.1982} can induce chiral-selective chemical reactions of chiral molecules (2-butanol) in an adsorbed adlayer due to the CISS effect. Later on, \cite{Rosenberg.2015} found that SPEs produced by X-ray irradiation on a nonmagnetic gold surface that are transmitted through a chiral overlayer\footnote{Here the chiral overlayer acts as a filter tha favors photoelectrons with a preferred spin state, so transmitted photoelectrons become spin-polarized.} could induce chiral-selective chemistry in an adsorbed adlayer, which is caused by the different quantum yields for the reaction of SPEs with two enantiomers (see \citealt{Rosenberg.2019} for a review). Using the chiral-induced selective spin (CISS) paradigm \citep{Ray.1999,Naaman.2012}, \cite{Ozturk.2022} suggested that SPEs induced by UV irradiation of magnetic deposits in the basin of an evaporative lake might induce CISS-driven reduction chemistry (CDRC) for prebiotic molecules in the lake-magnetite basin interface, resulting in the chiral asymmetry. Experimental study in \cite{Ozturk.2023a} showed that magnetic deposits act as chiral agents facilitating the homochiral enrichment of prebiotic compounds.

In \cite{Hoang.2024} (Paper I), we proposed that magnetically aligned dust grains play a key role in chiral asymmetry breaking for chiral molecules in astrophysical environments. First, aligned grains are a new source of SPEs due to the photoemission of electrons having aligned spins by the Barnett effect. Second, aligned grains induce chiral-dependent adsorption/crystalization of chiral molecules that would lead to chiral asymmetry of accumulated chiral molecules on aligned grains. They quantified the emission of SPEs from aligned grains due to the photoelectric effect of interstellar UV radiation. However, in dense molecular clouds and protostellar cores and disks, interstellar UV radiation is significantly attenuated. The dominant source of UV radiation is cosmic rays (CRs) because they can penetrate deep into dense regions. In particular, ice mantles of grains in dense clouds play a crucial role in astrochemistry, and COMs are thought to form in the grain ice mantle due to radical-radical reactions \citep{Herbst:2009go}. Polarization observations toward star-forming regions reveal that icy grains are efficiently aligned \citep{Whittet.2008,Vaillancourt:2020ch}. Moreover, irradiation of CR electrons and protons on aligned grains would also cause the emission of spin-polarized secondary electrons. Note that previous experiments demonstrated that irradiation of energetic electrons \citep{Unguris.1982} and ions \citep{Pfandzelter.2003} on ferromagnetic grains can produce SPE as when irradiated by UV/X-ray photons \citep{Kisker.1982}.

In this paper (Paper II), we explore the effect of CRs on the production of SPEs from aligned icy grains and their effect on chiral asymmetry of amino acids in dense protostellar environments where the UV interstellar radiation is substantially attenuated by high dust extinction. We focus on protostellar environments because these are the regions of active chemistry where most of COMs are detected. We study two important effects induced by CRs, including (1) photoemission of SPEs from aligned grains caused by UV radiation produced by the interaction of CRs and the gas, and (2) production of secondary SPEs by CRs bombardment onto aligned grains. 

The paper structure is described as follows. In Section \ref{sec:alignment}, we study the magnetic alignment of icy grains using the leading theory of grain alignment. In Section \ref{sec:CRs} we study the effects of CRs on the $\H_{2}$ ionization and resulting UV radiation. In Section \ref{sec:SPE_emission} we study the emission of SPEs by CR-induced UV radiation due to CR bombardment, and Section \ref{sec:SPE_production} is intended for discussion of the production of SPEs within aligned grains. Discussion and summary are shown in Sections \ref{sec:discuss} and \ref{sec:summary}, respectively.

\section{Magnetic Alignment of Icy Grains by Radiative Torques in Protostellar Cores}\label{sec:alignment}
Here we briefly describe the key elements of the leading theory for magnetic alignment of grains with magnetic fields based on radiative torques and derive the minimum size of aligned grains in dense regions. For a detailed discussion on grain alignment physics, please refer to our latest papers \citep{Hoang.2021,Hoang.2022}.

\subsection{Magnetic properties and Larmor precession of icy dust grains}
\begin{figure}
	\includegraphics[width=0.5\textwidth]{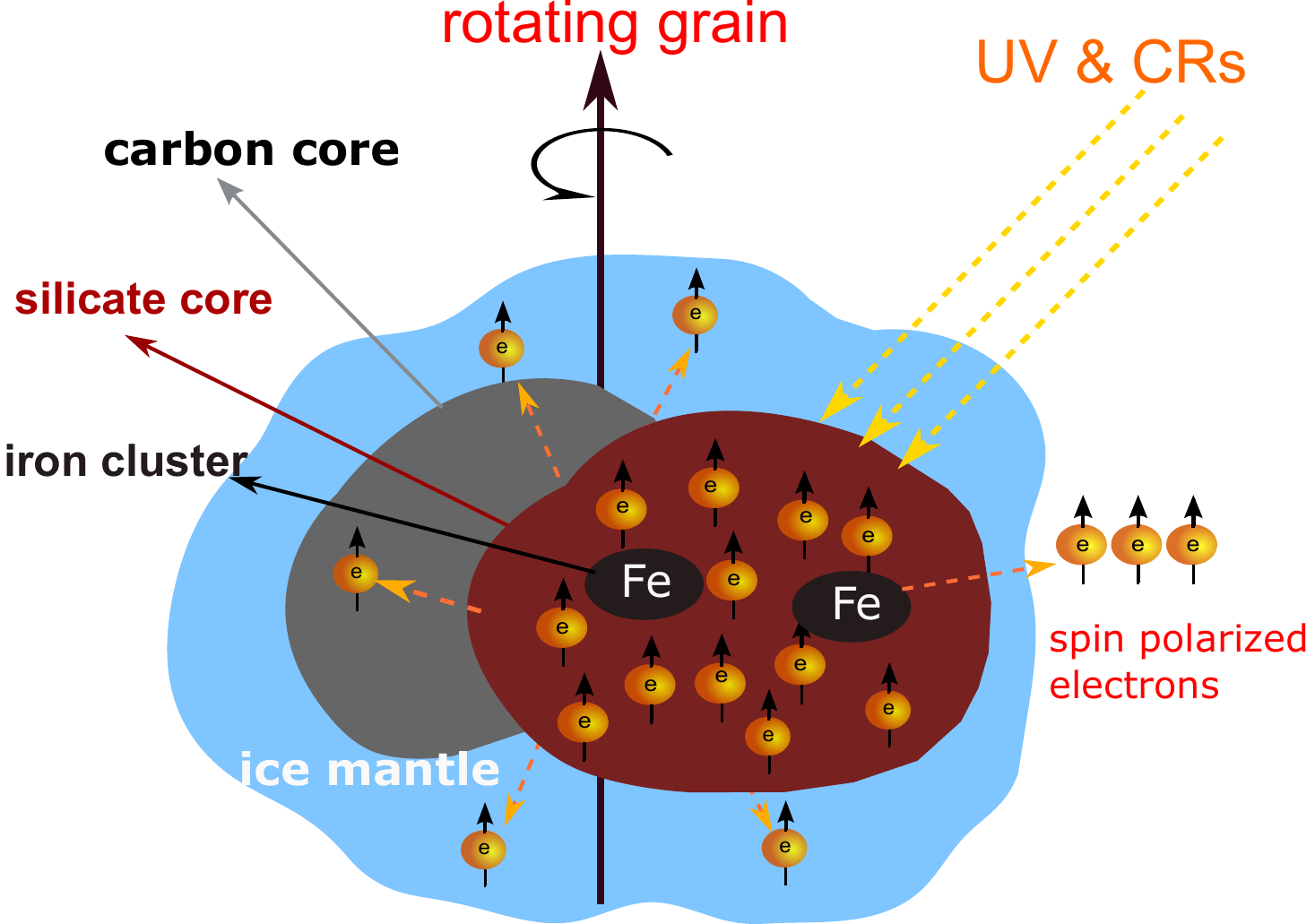}
	\caption{Schematic illustration of a model of icy grains in dense regions, including a grain core made of carbon and silicate core covered by an ice mantle. The silicate core contains embedded iron clusters. Irradiation of CRs and CR-induced UV radiation on the aligned rotating grain will produce spin-up or right-handed electrons.}
	\label{fig:icemantle_SPE}
\end{figure}

Magnetic properties of grains are essential for grain alignment with the magnetic field (\citealt{LazHoang.2007,HoangLaz.2016,LazHoang.2019}. To describe the magnetic properties of dust in dense regions, we first assume a feasible dust model for these environments, which includes a grain core equivalent to the Astrodust model suggested by \cite{Hensley.2021} for the diffuse ISM and an ice mantle as expected in dense and cold regions \citep{Greenberg.1998}. In the Astrodust model, the composite grain consists of a silicate core containing embedded iron clusters and a carbon core (see Figure \ref{fig:icemantle_SPE}). The volume of our icy grain model is the sum of the volumes of the silicate core ($V_{\rm sil}$), carbon core ($V_{\rm carb}$), and ice mantle ($V_{\rm ice}$):
\bea 
V_{\rm grain}&&=V_{\rm sil}+V_{\rm carb}+V_{\rm ice}\nonumber\\\
&&=V_{\rm sil}\left[1+(V_{\rm carb}/V_{\rm Sil}+(V_{\rm ice}/V_{\rm Sil})\right],\label{eq:Vgrain}
\ena
where $V_{\rm carb}/V_{\rm sil}\approx 0.36$ for the Astrodust model \citep{Hensley.2021}. The volume of the ice mantle is uncertain, but typically the mantle has a thickness of $L_{\rm ice}\sim 100\AA$ \citep{Greenberg.1998}. The effective radius of the icy grain is then $a=(3V_{\rm grain}/4\pi)^{1/3}$.

Silicate grains are assumed to contain embedded iron clusters as a result of grain coagulation in dense clouds, which are a superparamagnetic material. Therefore, the magnetic susceptibility of the silicate core can be given by 
\bea
\chi_{\rm sil,SPM}(0)\approx 0.52\left(\frac{N_{\rm cl}}{100}\right)\left(\frac{\phi_{\rm sp}}{0.1}\right)\left(\frac{p}{5.5}\right)^{2}\left(\frac{10\K}{T_{\rm d}}\right),\label{eq:chi_sp}
\ena
where $N_{\rm cl}$ is the number of iron atoms per cluster, $\phi_{\rm sp}$ is the volume filling factor of iron clusters within the silicate core, $p= \mu_{\rm at}/\mu_{B}$ with $\mu_{\rm at}$ being the atomic magnetic moment and $\mu_{B}=e\hbar/2m_{e}c$ Bohr magneton, and $T_{d}$ is the dust grain temperature (see \citep{HoangLaz.2016,HoangLaz.2016b}). Above, $N_{\rm cl}$ spans from $\sim 20$ to $10^{5}$, $\phi_{\rm sp}\sim 0.3$ if $100\% $ of Fe abundance present in iron clusters.  

For an icy grain rotating with angular velocity $\Omega$, the effective magnetic moment resulting from the Barnett effect \citep{Barnett.1915} (see e.g., \citealt{Hoang.2022}) is
\bea
\bmu_{\rm Bar}=-\frac{\bar{\chi}_{\rm grain}(0)V_{\rm grain}}{|\gamma_{e}|} \bOmega,
\label{eq:muBar}
\ena
where $\bar{\chi}_{\rm grain}(0)$ is the effective susceptibility of the icy grain of the volume $V_{\rm grain}$, and $|\gamma_{e}|\approx 2\mu_{B}/\hbar$ with $\mu_{B}=e\hbar/2m_{e}c$ is the Bohr magneton.

The total magnetic moment of the icy grain can be calculated as
\bea
\bmu_{Bar} &&= \bmu_{\rm sil} + \bmu_{\rm carb}+ \bmu_{\rm ice},\\
&& -\frac{\Omega}{|\gamma_{e}|}\left(\chi_{\rm sil,SPM}(0)V_{\rm sil}+\chi_{\rm carb}(0)V_{\rm carb}+\chi_{\rm ice}V_{\rm ice} \right),\label{eq:muBar_total}
\ena
where $\bmu_{j},\chi_{j}$ with $j=sil, carb, ice$ are the Barnett moment produced by the volume of silicate, carbon and ice mantle of their respective susceptibility $\chi_{j}$. 

The carbon core and water ice mantle can have a small paramagnetic susceptibility due to defect and hydrogen nuclear spins due to the lack of unpaired electrons \citep{Purcell.1979}. Here we set $\chi_{\rm carb}\sim \chi_{\rm ice}\sim 0$ due to their subdominance to the silicate core. Equaling Eqs. \ref{eq:muBar_total} to \ref{eq:muBar}, the effective magnetic susceptibility of icy grains is reduced by a factor of
\bea
\frac{\bar{\chi}_{\rm grain}(0)}{\chi_{\rm sil,SPM}(0)}&&=\frac{V_{\rm sil}}{V_{\rm grain}},\\
&&= \frac{1}{1+(V_{\rm carb}/V_{\rm sil}) +(V_{\rm ice}/V_{\rm sil})}.
\label{eq:chigrain_silcore}
\ena

For the typical thickness of ice mantle $L_{\rm ice}\sim 100\AA$, $V_{\rm ice}/V_{\rm sil}\sim 0.4$, so that the effective magnetic susceptibility is reduced by a small factor of $\bar{\chi}_{\rm grain}/\chi_{\rm sil,SPM}\sim 1/1.75\approx 0.57$ for the silicate core with a typical radius of $R_{\rm sil}=0.1\mum$. Therefore, the existence of thin ice mantles on the grain core does not significantly reduce the magnetic susceptibility of the superparamagnetic grain core. In realistic situations, Fe atoms from the gas phase can stick to the ice mantle, leading to Fe impurity and paramagnetic property of the ice mantle. This picture is supported by observations showing the gradual of Fe abundance in the gas phase from the diffuse ISM to MC \citep{Savage.1979,Zhukovska:2008p3096,Jenkins.2009}. Indeed, \cite{Sorrell.1994} suggested that Fe impurity, even of only 1\% of Fe abundance in the ice mantle, can be transformed into iron oxide (Fe$_3$O$_4$) due to heating by grain-grain collisions. Moreover, the bombardment of low-energy CRs can create a track in the ice and facilitate the aggregation of Fe$_3$O$_4$ molecules into a cluster. This mechanism will form superparamagnetic maghetite clusters and make the ice mantle become superparamagnetic material. This is a promising process for enhancing the magnetic susceptibility of ice grains in dense regions.

The interaction of the grain magnetic moment (Eq. \ref{eq:muBar}) with the external magnetic field causes the regular (Larmor) precession of the grain angular momentum around the magnetic field direction. The characteristic timescale of such a Larmor precession is given by
\bea
\tau_{\rm Lar}&=&\frac{2\pi I \Omega}{|\mu_{\Bar}|B}=\frac{2\pi |\gamma_{e}|I}{\bar{\chi}_{\rm grain}(0)V_{\rm grain}B},\nonumber\\
&\simeq &8.4\times 10^{-4}\hat{\rho} a_{-5}^{2}\left(\frac{0.1}{\bar{\chi}_{\rm grain}}\right)\left(\frac{5\mu G}{B}\right)~\yr,
\label{eq:tauB}
\ena
where $I=8\pi \rho a^{5}/15$ is the grain inertia moment, $\rho$ is the grain mass density and $\hat{\rho}=\rho/(3\g\cm^{-3})$, and $B$ is the magnetic field strength of the cloud \citep{Hoang.2022}. The Larmor precession of grains with embedded iron clusters is much faster than the randomization of grain orientations by gas random collisions in protostellar environments \citep{Hoang.2021,Hoang.2022}, so that the magnetic field is the axis of grain alignment.

\subsection{Grain alignment with magnetic fields by radiative torques}
Magnetic dust grains can align with the ambient magnetic field due to different physical processes, including the internal alignment of the grain axis of maximum inertia (e.g., short axis) with the angular momentum and the external alignment of the grain angular momentum with B-fields \citep{Lazarian.2007,LAH.2015,Hoangetal.2022}. The internal alignment is caused by Barnett and inelastic relaxation \citep{Purcell.1979}, which is found to be efficient for grains with embedded iron inclusions \citep{Hoang.2021,Hoang.2022}. The external alignment is governed by radiative torques \citep{LazHoang.2007} and enhanced paramagnetic relaxation by magnetically enhanced radiative torque (MRAT) mechanism \citep{LazHoang.2008,HoangLaz.2008,HoangLaz.2016}.

Numerical simulations in \cite{HoangLaz.2008,HoangLaz.2016} show that if the RAT alignment has a high-J attractor point, then, large grains can be perfectly aligned because grains at low-J attractors would be randomized by gas collisions and eventually transported to more stable high-J attractors by RATs. Here, the low-J and high-J attractors are defined as the attractor at which the grain has the angular velocity comparable and much larger than its thermal angular velocity, $\Omega_{T}=\left(2kT_{\rm gas}/I\right)^{1/2}$ with $T_{\rm gas}$ being the gas temperature. On the other hand, grain shapes with low-J attractors would have negligible alignment due to gas randomization. For small grains, numerical simulations show that the alignment degree is rather small even in the presence of iron inclusions because grains rotate subthermally \citep{HoangLaz.2016}. Therefore, the degree of grain alignment depends critically on the critical size above which grains can be aligned by RATs, denoted by $a_{\rm align}$, which is determined by the grain size at which its rotation rate spunup by RATs, $\Omega_{\rm RAT}= 3\Omega_{T}$.

Using the RAT paradigm, the minimum size of grain alignment depends on the local conditions of the gas and radiation field. For the dense core without embedded stars, grains are aligned by attenuated diffuse interstellar radiation field (ISRF). Following \cite{Hoang.2021}, the alignment size increases with the visual extinction $A_{\rm V}$ measured from the cloud surface inward as
\bea
    a_{\rm align} \simeq &&  0.11\hat{\rho}^{-1/7} \left(\frac{\gamma_{-1} U_{0}}{n_{4}T_{0,1}}\right)^{-2/7}\left(\frac{\bar{\lambda}_{0}}{1.2\mum}\right)^{4/7}\nonumber\\
&&\times (1+0.42A_{\rm V}^{1.22})^{(2-2/(4+\beta))/7} \nonumber\\
&&\times (1+0.27A_{\rm V}^{0.76})^{4/7} \mum,
\label{eq:aalign_AV_GMC}
\ena
where $n_{4}=n_{\H}/(10^{4}\cm^{-3})$ with $n_{\H}$ is the normalized nucleon density, $\bar{\lambda}_{0}$ and $\gamma=0.1\gamma_{-1}$ are the mean wavelength and anisotropy degree of the ISRF, $T_{0,1}=T_{0}/10\K$ with $T_{0}\sim 15\K$ is the dust temperature at the surface of the cloud irradiated by the ISRF of strength $U_{0}=u_{\rm ISRF}/u_{\rm MMP}\sim 1$ with $u_{\rm MMP}=8.64\times 10^{-13}\erg\cm^{-3}$ the energy density of the ISRF in the solar neighborhood from \cite{Mathis.1983}, and $\beta\sim 1.5$ is the dust spectral index in dense cores.


Using the above equation, we can estimate the alignment size for three typical locations of the dense core (outer, inner, and central) considered in \cite{Ivlev.2015a}, with the typical volume ($n_{\H_{2}}$) and column ($N_{\H_{2}}$) densities of $[n_{\H_{2}}, N_{\H_{2}}]=[10^{4}\cm^{-3}, 3.2\times 10^{21}\cm^{-2}, [10^{6}\cm^{-3}, 2.8\times 10^{22}\cm^{-2}], [2\times 10^{7}\cm^{-3},10^{23}\cm^{-2}]$. The corresponding values of visual extinctions are calculated using the typical formula $A_{\rm V}=(N_{\H_{2}}/5.1\times 10^{21}\cm^{2})\times (3.1/R_{\rm V})$, which yields $A_{\rm V}=[1, 11, 39]$. The alignment size at these three locations are calculated using Equation \ref{eq:aalign_AV_GMC} using $n_{\H}\approx 2n_{\H_{2}}$. The results are shown in Table \ref{tab:densecores}. One can see that large grains of size $a\gtrsim 1\mum$ are still well aligned up to the inner region of the core. However, only very large grains of $a>5\mum$ can be aligned at the central region of the core due to the significant attennuation of the ISRF and the increase in the gas density (see \citealt{Hoang.2021} for more details).

\begin{table}
	\caption{The minimum size of grain alignment by RATs in a starless core}
	\begin{tabular}{l|l|l|l}
		Parameters & Outer & Inner & Center\cr
		\hline\hline\cr
		N$_{\H_{2}}$[$\cm^{-2}$] & $3.2\times 10^{21}$ & $2.8\times 10^{22}$ & $10^{23}$\cr
		n$_{\H_{2}}$[$\cm^{-3}$]& $10^{4}$ & $10^{6}$ & $2\times 10^{7}$ \cr
		$A_{\rm V}$[mag] & 1 & 11 & 39 \cr
		$a_{\rm align}(\mum)$ & 0.12 & 0.99 & 4.9\cr
	\end{tabular}
	\label{tab:densecores}
\end{table}

In the presence of an embedded protostar, grain alignment can be driven by attennuated protostellar radiation. Consider a protostellar core with the density profile $n_{\H}=n_{c}\sim 10^{7}-10^{8}\cm^{-3}$ for $r<r_{c}$ and $n_{\H}=n_{c}(r/r_{c})^{-p}$ with $p\sim 1.5-2$ for $r>r_{c}$ where $r_{c}$ is the radius of the central region of $\sim 30\AU$ for low-mass and $r_{c}\sim 500\AU$ for high-mass protostars. Let $L_{\star}$ and $T_{\star}$ be the bolometric luminosity and effective temperature of the protostar. The typical visual extinction of the central region is $A_{\rm V}^{*}(r_{c})\sim 30$ for low-mass protostars and $A_{\rm V}^{*}(r_{c})\sim 50$ for high-mass protostars (see \citealt{Hoang.2021}). Therefore, the UV protostellar radiation is significantly attennuated beyond a small region of $A_{\rm V}^{*}\sim 10$ around the protostar. Thus, the major volume of the protostellar core is dominantly irradiated by optical-NIR photons, with the mean wavelength of the protostellar radiation increasing rapidly with $A_{\rm V}^{*}$ \citep{Hoang.2021}. The attenuated protostellar radiation, although has negligible effect on photoelectric effect, is essential for driving grain alignment by RATs. Following \citep{Hoang.2021}, the alignment size at radius $r$ and a visual extinction $A_{\rm V}^{*}$ measured from the protostar is estimated by 
\bea
a_{\rm align}
&\simeq &0.031\hat{\rho}^{-1/7} \left(\frac{U_{\rm c,6}}{n_{\rm c,8}T_{\rm in,2}}\right)^{-2/7}\left(\frac{\bar{\lambda}_{\star}}{1.2\mum}\right)^{4/7}  \nonumber\\
&&\times (1+c_{1}A_{V,\star}^{c_{2}})^{(2-q)/7}  (1+c_{3}A_{V,\star}^{c_{4}})^{4/7}\nonumber\\
&&\times \left(\frac{r}{r_{\rm c}}\right)^{2(2-p-q)/7} (1+F_{\rm IR})^{2/7}\mum,
\label{eq:aalign_AV_star}
\ena
where $F_{\rm IR}$ is the dimensionless parameter for describing the rotational damping by infrared dust emission, $U_{c,6}=U_{c}/10^{6}$ where $U_{c}=u_{\rm rad,c}/u_{\rm MMP}$ with $u_{\rm rad,c}=L_{\star}/4\pi r_{c}^{2}c$, $q=2/(\beta+4)$, and $c_{1}-c_{4}$ are the fitting parameters for the attenuation of protostellar radiation field (see Table 2 in \citealt{Hoang.2021}). 

\begin{figure}
    \includegraphics[width=0.5\textwidth]{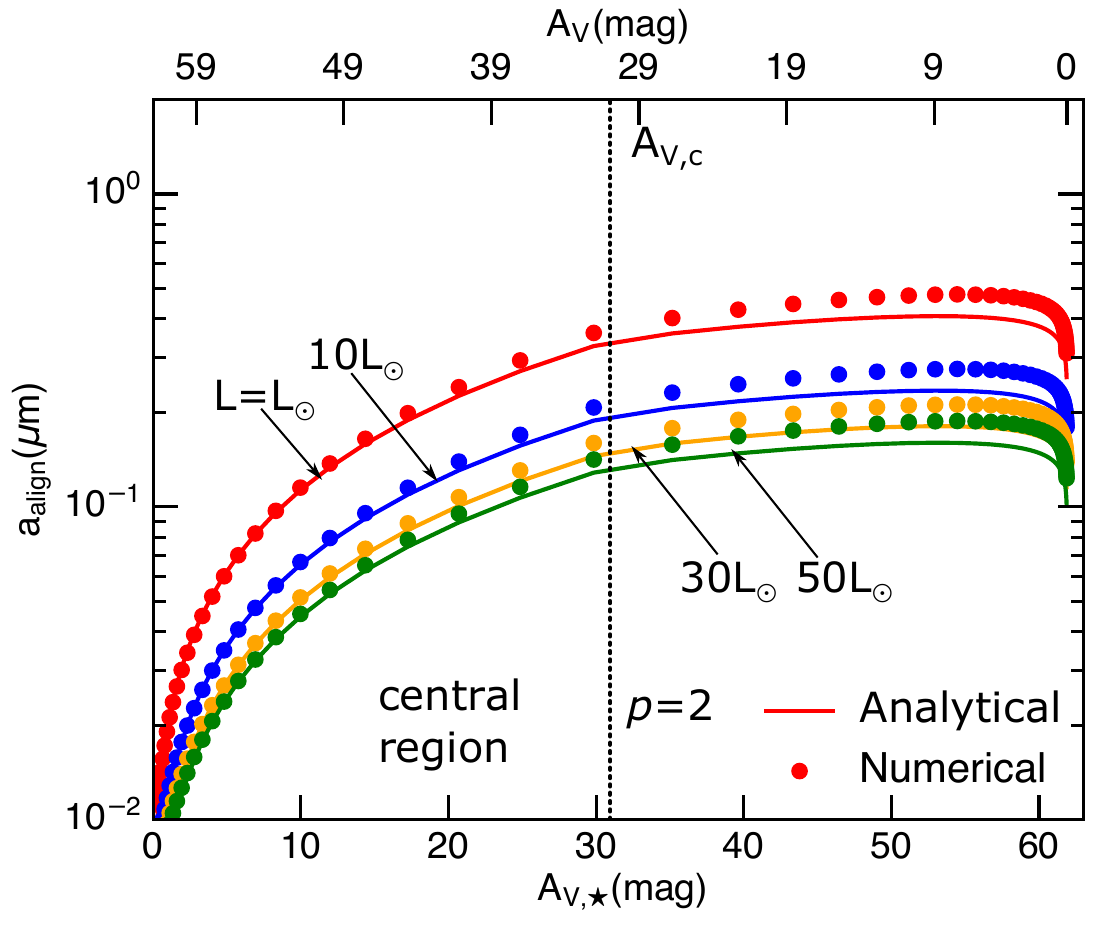}
    \caption{The minimum size of grain alignment by RATs in a protostellar core with the density profile with the slope of $p=2$ for the different luminosities. The vertical line marks the boundary between the central region and the envelope with $A_{V,\star}\sim 30$. The alignment size increases with $A_{V,\star}$ due to attennuation of protostellar radiation. Results from \cite{Hoang.2021}.}
    \label{fig:alignment_protostar}
\end{figure}

Figure \ref{fig:alignment_protostar} shows the alignment size $a_{\rm align}$ for the protostellar core with the embedded protostars of different luminosity. For the low-mass protostar of $L_{\star}=1L_{\odot}$ and $T_{\star}=3000\K$, small grains larger than $a>a_{\rm align} \sim 0.1\mum$ can be aligned by protostellar radiation at the central region of $A_{V,\star}<30$, but only large grains of $a\sim 0.5\mum$ can be aligned in the outer regions of $A_{V,\star}\sim 50-60$. For a high-mass protostar of $L_{\star}\sim 10^{3}L_{\odot}$ and $T_{\star}=10^{4}\K$, the alignment sizes are smaller (see Table \ref{tab:densecores}; for more details see Fig. 8 of \citealt{Hoang.2021}). Note that observations reveal significant grain growth in dense protostellar environments and very large grains of size $a>10\mum$ can be present (see \citep{Kwon.2019,Miotello.2014,Galametz:2019fj}). Therefore, large grains of $a\gtrsim 0.1\mum$ in the protostellar cores can be magnetically aligned by RATs, as numerically modeled in \citep{GiangHoang.2023,GiangHoang.2024}.

\section{Cosmic Rays and and Cosmic Rays-Induced UV Radiation in Protostellar Cores}
\label{sec:CRs}
\subsection{Cosmic Ray Transport and CR Spectrum}
Cosmic rays can penetrate into dense regions of the ISM and are expected to play an important role in producing SPEs in protostellar cores. CRs can produce SPEs either directly in collisions with aligned grains (by collisional ionization) or indirectly via the photoelectric effect by CR-induced UV radiation. To estimate the production rate of SPEs in protostellar cores, we need to know first the interstellar spectra of Galactic CRs and then perform the modeling of CRs transport in the dense regions to obtain attennuated CRs spectra.   

The interstellar spectra of CRs, are actually not very well known, especially in the low-energy ($\lesssim1$ MeV) range which, as we shall see later, is particularly relevant for SPE production. This is partially due to the fact that the main sources of Galactic CRs are not yet unambiguously identified. Many different classes of sources have been proposed for high-energy CRs (from GeV to PeV), e.g supernova remnants (SNRs, see e.g. \citealt{gabici2019,cristofari2021b} for critical reviews), superbubbles \citep{parizot2004,vieu2022}, or star cluster \citep{morlino2021}. Recently, progress on low-energy CRs has also been made thanks to the availability of data from Voyager probes and measurements of CR ionization rates\footnote{Here, the CR ionization rate is defined as the production rate of H$_2^+$ ions per H$_2$ molecules \citep[see e.g.][]{neufeld2017}.} via molecular line observations. Some new classes of sources for MeV CRs have also been put forward in the literature for example, H\rom{2} regions \citep{meng2019,padovani2019}, protostellar jets \citep{padovani2015,padovani2016,gaches2018}, or wind termination shocks of stars \citep{scherer2008}. Interestingly, enhanced CR ionization rates have been observed around several SNRs \citep{vaupre2014,gabici2015,phan2020} which strongly indicate that these objects can be responsible for the entire Galactic CR spectra from MeV to PeV and maybe beyond. The expected interstellar spectra of CRs from SNRs match quite well with local observations of the CR spectra \citep[see e.g.][]{phan2021}. 

For the sake of convenience and also account for the uncertainty in observed CR spectra, we adopt a parametric model for the Galactic CR spectra as in \cite{Ivlev.2015a}, which is described by
\bea
j_{s}(E)=C\frac{E^{\alpha}}{(E+E_{0})^{\beta}}~\eV^{-1}\cm^{-2}\s^{-1}\sr^{-1},\label{eq:jCR_ISM}
\ena
where $s=e, p$ denotes the CR electrons and protons, $\alpha$ and $\beta$ are the model parameters, $E_{0}=500$ MeV is the turnover energy of CR spectrum, and $C$ is the normalization constant. For CR electrons, $\alpha=-1.5, \beta=1.7$, and $C=2.1\times 10^{18}$. For CR protons, we consider two models, model L (low) and model H (high), which have the parameters $\alpha=0.1,\beta=2.8$ (model L) and $\alpha=-0.8,\beta=1.9$ (model H) with $C=2.4\times 10^{15}$. We assume the lower cutoff of the spectrum to $E_{\rm cut}=1$ keV and upper cutoff of $E_{\rm max}=10$ GeV.

The Galactic CRs undergo energy loss due to interaction with the gas when penetrating into the dense core. This results in the attenuation of CRs with increasing the depth into the cloud or the column density $N_{\H_{2}}$. To calculate the attenuated CR spectrum in the protostellar core at the column density $N\equiv N_{\H_{2}}$, we follow the continous slowing down approximation (CDSA) model (see \citealt{Padovani.2009}), which yields
\bea
j_{k}(E,N)=j_{k}(E_{0})\frac{L_{k}(E_{0})}{L_{k}(E)},\label{eq:jCR_NH}
\ena
where $k=e,p$, $E_{0}$ is the initial energy and $E$ is the energy of CRs that has passed a column density $N_{\rm H}$, $L_{k}(E_{0})$ and $L_{k}(E)$ are the energy loss function at $E_{0}$ and $E$ respectively. 

\begin{figure}
	\centering
	\includegraphics[width=0.5\textwidth]{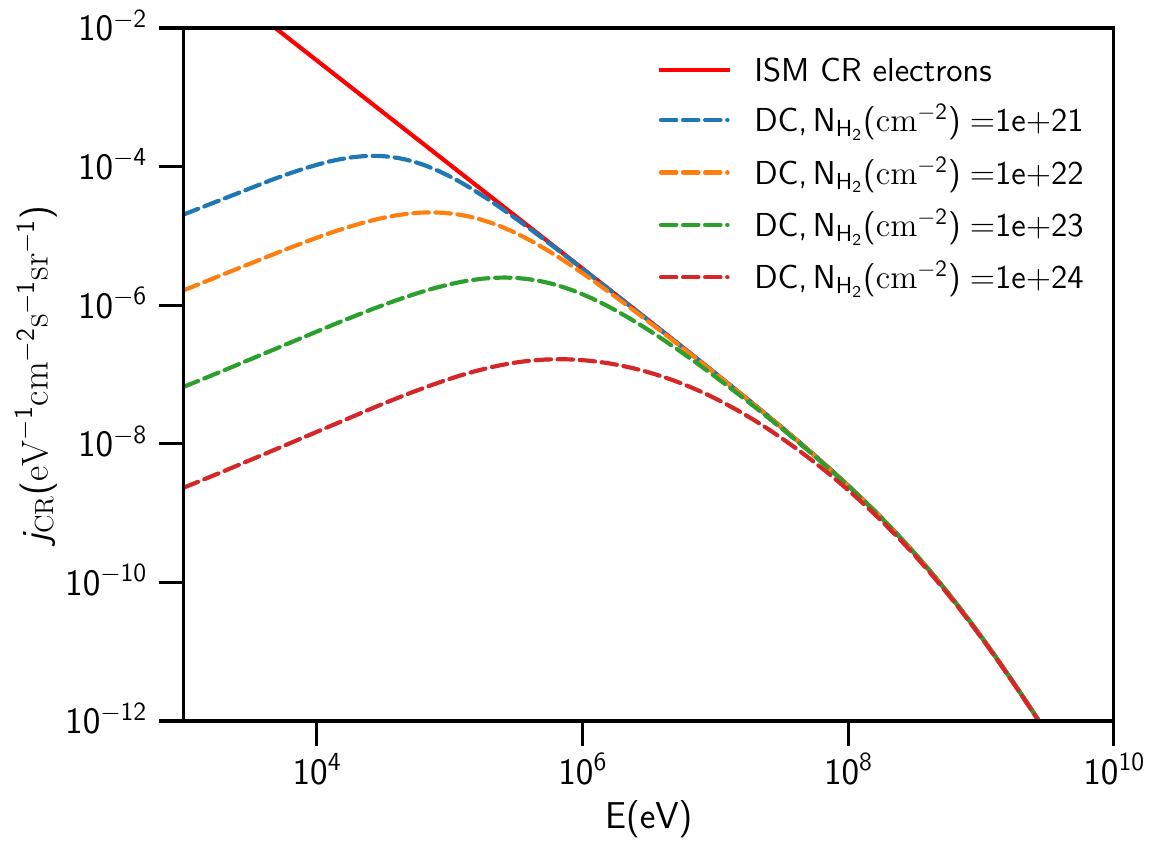}
	\includegraphics[width=0.5\textwidth]{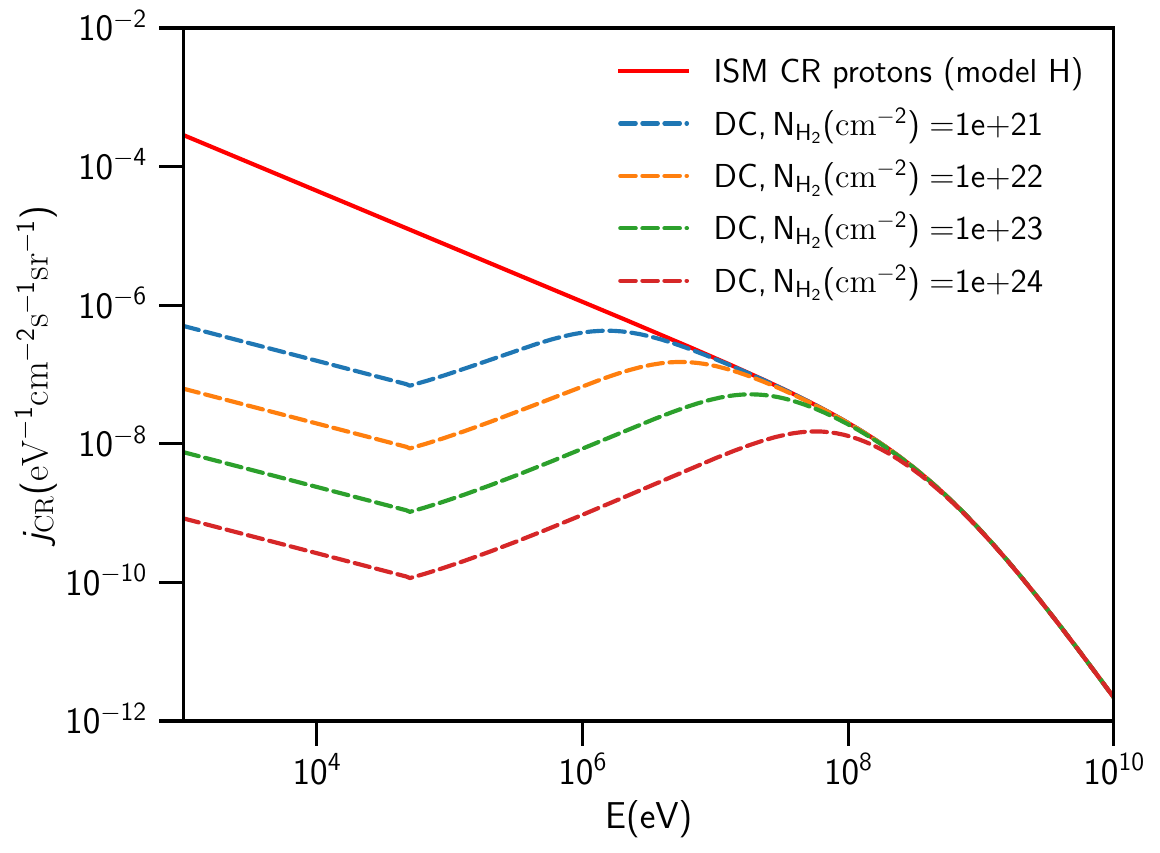}
	\includegraphics[width=0.5\textwidth]{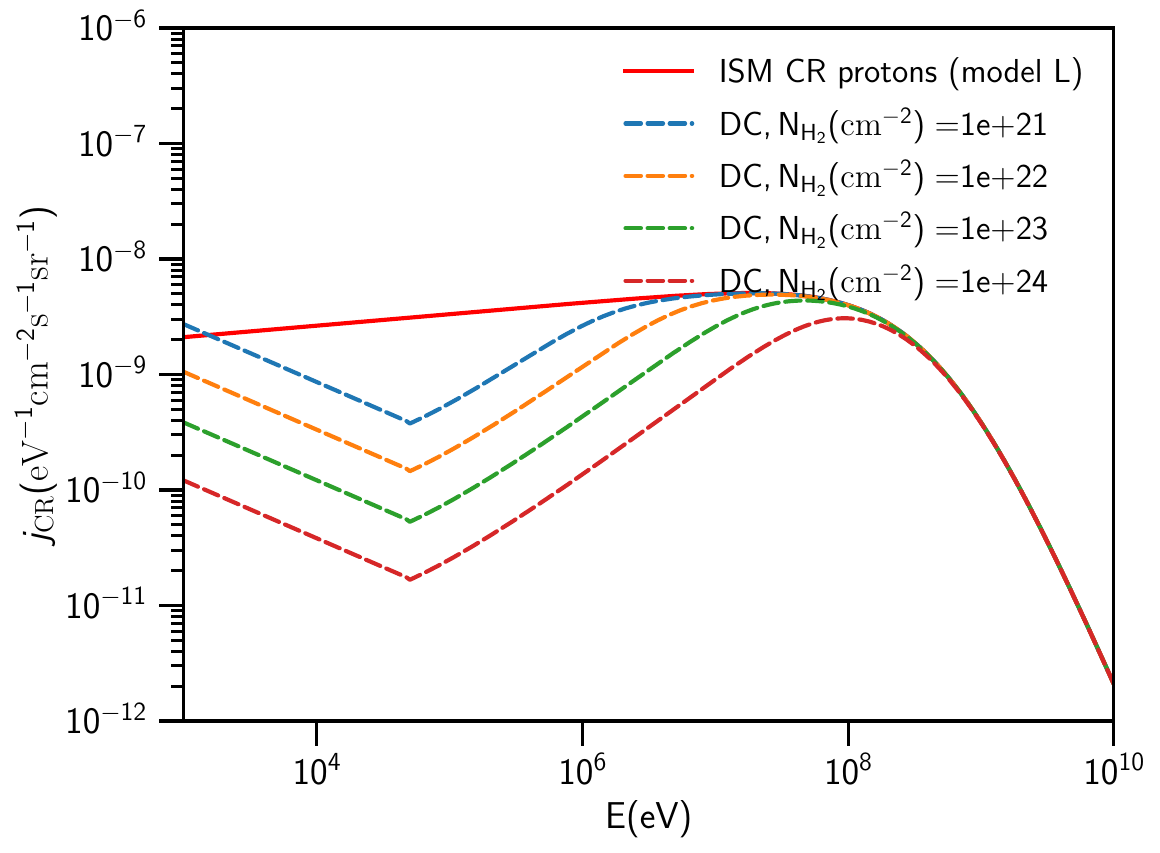}
	\caption{Spectra of CR electrons (top panel) and CR protons (middled and bottom panels for model H and L) at the diffuse ISM (solid lines) and estimated in dense clouds (DC) with different column densities $N_{\H_{2}}$ using the continuous slowing down approximation model.}
	\label{fig:jCR}
\end{figure}

A detailed description of the CDSA model and numerical calculations are shown in Appendix \ref{apdx:CRtransport}. In Figure \ref{fig:jCR}, we only show the final results of the spectra of attenuated CR electrons and protons for the different column densities. The flux of low-energy CRs of energy below $1$ MeV is significantly decreased with increasing $N_{\H_{2}}$ due to the ionization and electronic excitations of $\H_{2}$. The high-energy CRs of $E>100$ MeV are hardly attenuatted.


\begin{figure}
    \centering
    \includegraphics[width=0.5\textwidth]{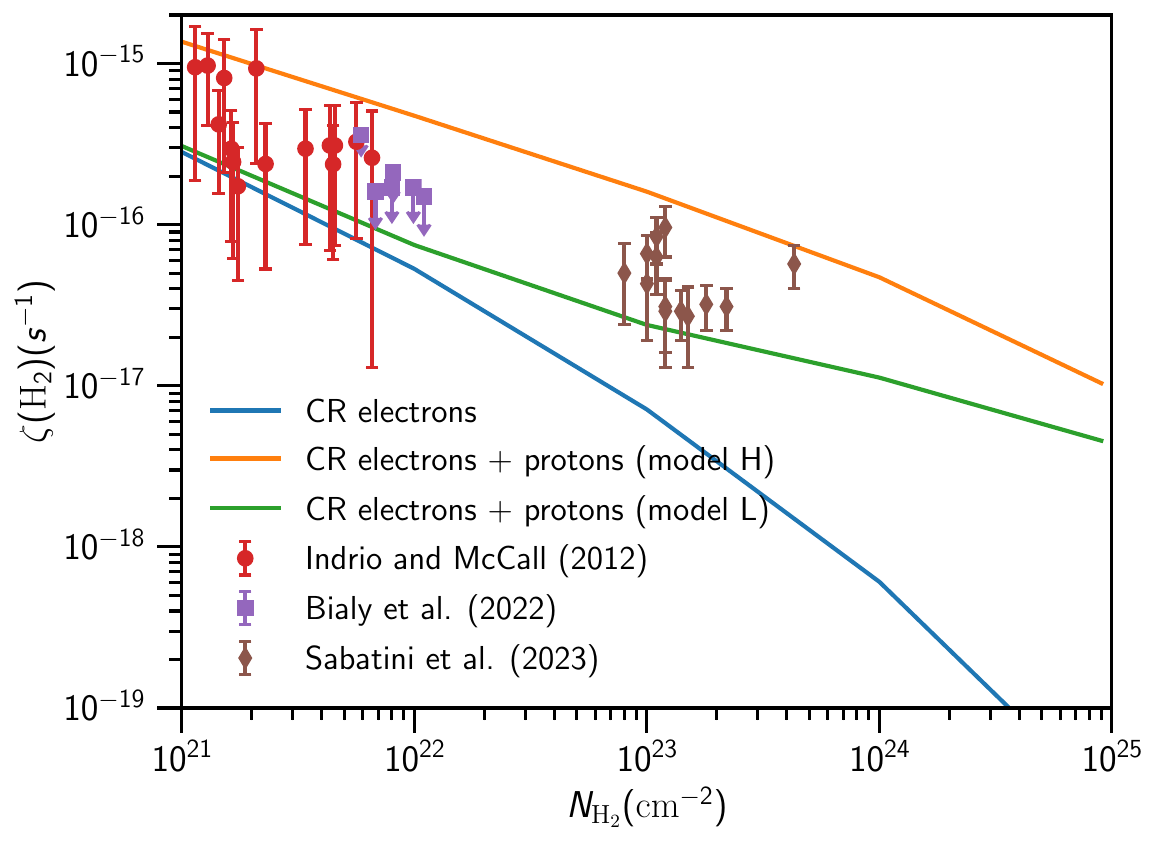}
    \caption{Cosmic-ray ionization rates calculated for the different column densities and comparison to observational data from \citet{indriolo2012} and \citet{sabatini2023}, and upper limits from \citet{bialy2022}. The L and H models constitute the lower and upper bounds for the observations above $3\times 10^{21}\cm^{-2}$.}
    \label{fig:Zeta_CR}
\end{figure}

\subsection{CR ionization rate per $\H_2$}
The total CR ionization rate per H$_2$ can be calculated as (see \citealt{Padovani.2009})
\bea
\zeta(\H_{2})&&= \int_{I(\H_{2})}^{E_{\rm max}} (1+\phi(E))\sigma_{\H_{2}}^{\rm CRe}(E)4\pi j_{\rm CRe}dE\nonumber\\
&&+\int_{I(\H_{2})}^{E_{\rm max}} (1+\phi(E))\sigma_{\H_{2}}^{\rm CRp}(E)4\pi j_{\rm CRp}dE\nonumber\\
&&+ \int_{0}^{E_{\max}}\sigma_{\H_{2}}^{\rm CRp,e.c} j_{\rm CRp}dE,
\label{eq:zeta_H2}
\ena
where $I(\H_{2})=15.603\eV$, $\phi(E)$ is the correction factor accounting for ionization of H$_2$ by secondary electrons, $\sigma^{\rm CRe},\sigma^{\rm CRp}$ are the ionization cross-sections of $H_{2}$ by CRe and CRp, and $\sigma^{\rm CRp,ec}$ is the cross-section for CR proton capture by $\H_2$ via the process $p + \H_{2}\rightarrow \H + \H_{2}^{+}$ (see Appendix \ref{apdx:cross-section}).

Using the attennuated CR spectra from Figure \ref{fig:jCR}, we calculate $\zeta(\H_{2})$ by both CR electrons and protons. Results are shown in Figure \ref{fig:Zeta_CR} as a function of the column density $N_{\H_{2}}$ (solid line). Measurements from various observations of dense clouds and protostellar cores \citep{caselli1998,indriolo2012,redaelli2021,bialy2022,sabatini2023} are overplotted for comparison. The data show strong scatters in the diffuse region of $N_{\H_{2}}<3\times 10^{21}\cm^{-2}$, but can be well constrained by the model L and model H for dense regions of $N_{\H_{2}}>3\times 10^{21}\cm^{-2}$. The CR ionization tends to decrease with increasing the gas column density due to the CR energy loss.

\subsection{UV radiation from H$_2$ fluorescence by cosmic rays}
 The interaction of CRs with interstellar gas can produce radiation via several mechanisms, including CR electron Bremsstrahlung, H$_2$ fluorescence, and $\pi^{0}$ decay (see e.g., \citealt{Ivlev.2015a}). The $H_2$ fluorescence produces UV photons with energy around 11.2-13.6 eV due to CR-induced excitations of H$_2$ molecules into the Lyman-Werner states \citep{Ivlev.2015a}, which is the most important process for the production of SPEs via photoelectric effect.

The flux of CR-induced UV photon flux $\Phi_{\rm UV}$ can be approximately estimated for high-column-density clouds using the CR ionization rate $\zeta({\H}_2)$ by \citep{Padovani.2024}
\bea
\Phi_{\rm UV}\simeq 
100(c_{0}+c_{1}R_{\rm V}+c_{2}R_{\rm V}^{2})\left(\frac{\zeta({\H}_2)}{10^{-17}\,{\rm s}^{-1}}\right)~\cm^{-2}\s^{-1}.\label{eq:PHIUV_CR}
\ena
where $[c_{0},c_{1},c_{2}]=[5.023, -0.504, 0.115]$, $R_{V}$ is total-to-selective dust extinction. The above linear relation is only valid for $3.1\le R_{\rm V}\le 5.5$ and in the dense regions with $N_{\rm H}\gtrsim 10^{22}\cm^{-2}$. 

Also, since CR-induced UV radiations are produced by secondary electrons from CR ionization events, a linear relation between $\Phi_{\rm UV}$ and $\zeta({\rm H}_2)$ is expected. Using the spectra of attenuated CRs obtained in the previous section (Fig. \ref{fig:jCR}), we can calculate $\Phi_{\rm UV}$ using Equation \ref{eq:PHIUV_CR}. 

\subsection{Thermal Electrons produced by CR ionization of $\H_{2}$}
In dense regions, the main source of gas ionization is by CRs and CR-induced UV radiation. The gas ionization fraction, $x_{e}$, can be described by \citep{Ivlev.2015a}
 \bea
 x_{e}\simeq 1.7\times 10^{-8}\left(\frac{n_{\H_{2}}}{10^{4}\cm^{-3}}\right)^{-0.65}\left(\frac{\zeta(\H_{2})}{10^{-17}\s^{-1}}\right)^{1/2}, \label{eq:xe}
 \ena
 which increases with the CR ionization rate and decreases with the gas volume density.

The number density of thermal electrons is then calculated by $n_{e}=x_{e}n(\H_{2})$, which yields the flux of thermal electrons
\bea
\Phi_{\rm Te}&&=x_{e}n_{\H_{2}}v_{\rm Te},\nonumber\\
&&\simeq 3642\left(\frac{n_{\H_{2}}}{10^{4}\cm^{-3}}\right)^{-0.35}\left(\frac{T_{\rm gas}}{100\K}\right)^{1/2}\nonumber\\
&&\times
\left(\frac{\zeta(\H_{2})}{10^{-17}\s^{-1}}\right)^{1/2}.\label{eq:j_Te}
\ena

Compared to the flux of CR electrons, one can see that $\Phi_{\rm Te}$ is two orders of magnitude higher than the CR electron flux. However, $\Phi_{\rm Te}$ is slightly higher than the UV flux (see Eq. \ref{eq:PHIUV_CR}). 

\section{Spin Polarization and Emission of SPEs from Aligned Grains}
\label{sec:SPE_emission}
CR electrons and thermal electrons produced by $H_2$ ionization by CRs are expected to have random spins, i.e., un-polarized. Here, we first show that the collection of thermal and CR electrons/protons onto aligned grains will quickly produce spin-polarization due to the Barnett effect. We then discuss the emission of SPEs by CR-induced energetic radiation and CR electrons.

\subsection{Collection of thermal electrons and CR electrons}
\subsubsection{Thermal electrons}
Thermal electrons will be collected by the dust grain upon collisions. The collection rate of thermal electrons by a grain of size $a$ is given by
\bea
J_{\rm coll,Te} &=&n_{e}v_{\rm Te}\pi a^{2}g(\phi_{Z})~\s^{-1},\nonumber\\
&&\simeq 5 \times 10^{-7}a_{-5}^{2}n_{5}^{-0.35}g(\phi_{Z})\left(\frac{T_{\rm gas}}{100\K}\right)^{1/2}\nonumber\\
&&\times
\left(\frac{\zeta(\H_{2})}{10^{-17}\s^{-1}}\right)^{1/2}
\label{eq:Jcoll_e}
\ena
where $v_{\rm Te}=(8kT_{\rm gas}/(\pi m_{e}))^{1/2}$ is the thermal speed of electrons, $g(\phi_{Z})$ with $\phi_{Z}=Z_{\rm grain}e^{2}/a$ and $Z_{\rm grain}$ the grain charge. Here, $g =e^{\phi_{Z}/kT_{\rm gas}}$ for $Z_{\rm grain}<0$ and $g(\phi_{Z})=1+\phi_{Z}/kT_{\rm gas}$ for $Z_{\rm grain}\ge 0$ that accounts for the Coulomb effect on electron collision cross-section. Throughout this explorary paper, we consider neutral grains of $Z_{\rm grain}=0$ for convenience.


\subsubsection{Collection of CR Electrons}
Grains may also capture CR electrons and protons upon CR bombardment. The collection rate of CR electrons by a grain of size $a$ is given by \cite{Draine.1979}
\bea
J_{\rm coll,CRe}(a,Z)=\pi a^{2}\int_{E_{int}}^{\infty} 4\pi j_{\rm CRe}(E) s_{e}(E) dE,\s^{-1},
\ena
where $s_{e}$ is the sticking coefficient of CR electrons, and $E_{\rm int}$ is the minimum energy for the grain-CR interaction. 

The sticking coefficient $s_{e}(E)$ is equal 1 for the stopping range $R_{e}(E)<4a/3$ and $s_{e}=0$ for $R_{e}(E)>4a/3$ where
\bea 
R_{e}(E)= (An)^{-1}E^{n}\simeq 0.03\rho^{-0.85}(E/1{\rm keV})^{n}\mum,\label{eq:Re}
\ena
where $n = 1.5$ for electron energy from 300 eV to 1 MeV (\citealt{Draine.1979}, see \citealt{Hoang.2015}). One can see that for CR electrons of $E>100$ keV, the stopping range is $R_{e}(E)>30\mum$. Therefore, for grains of $a\sim 1\mum$ in protostellar environments, CR electrons do not stick to the grain but pass through, so one has $s_{e}=0$. The same effect is expected for CR protons.

Since the flux of CR electrons and protons, $j_{\rm CRe, CRp}$, is much lower than the flux of thermal electron density $j_{\rm Te}$ (see Eq. \ref{eq:j_Te}), the collection of CR electrons/protons is negligible compared to the collection of thermal electrons. 

\subsection{Spin Polarization of Captured Electrons by the Barnett Effect}
Electrons captured by a dust grain are most likely present in the outermost shell of atomic energy configuration and become unpaired electrons of the grain atoms. These unpaired electrons with initially random spins will be polarized due to the Barnett effects and have spins aligned along the grain rotation axis if the timescale for spin alignment is much shorter than the time interval between two CR electron bombardments. Let us estimate these timescales for dust grains.

The equivalent magnetic field induced by the Barnett effect for a rotating grain of angular velocity $\Omega$ is given (e.g., \citealt{HoangBao.2024})
\bea
B_{\rm Barnett}&=&\frac{\Omega}{|\gamma_{e}|}=\frac{2m_{e}c}{eg_{e}}\Omega,\nonumber\\
&\simeq & 3\hat{\rho}^{-1/2}T_{g,1}^{1/2}s^{-1/2}a_{-5}^{-5/2} \left(\frac{\Omega}{\Omega_{T}}\right)~{\rm mG},\label{eq:Beq}
\ena
 where $g_{e}\approx 2$ for electrons, and $s$ is the parameter describing grain shape of the order of unity. For aligned grains of $a>a_{\rm align}$ by RATs, one has $\Omega/\Omega_{T}\gtrsim 3$. Thus, aligned grains would produce $B_{\rm Barnett}\gtrsim 1$mG.
 
The Larmor precession timescale of the electron spin in this Barnett magnetic field is given by
\bea
\tau_{\rm Lar}&& = \frac{2\pi }{\omega_{\rm Lar}}=\frac{2\pi S}{\mu_{el}B_{\rm Barnett}}\\
&&\simeq \frac{3.5\pi m_{e}c}{eB_{\rm Barnett}} \simeq 6.2\times 10^{-4}\frac{1mG}{B_{\rm Barnett}}\s,\label{eq:tauLar}
\ena
where $S=\sqrt{s_{e}(s_{e}+1)}\hbar$ with $s_{e}=1/2$ for electrons.

For the flux of CR electrons of $\Phi_{\rm CRe}=n_{\rm CRe}\bar{\rm V}_{\rm CRe}$ with $\bar{\rm V}_{\rm CRe}$ the average speed of CR electrons, the time interval between two CRe bombardments is 
\bea
\tau_{\rm CRe}&=&\frac{1}{\pi a^{2}n_{\rm CRe}\bar{\rm V}_{\rm CRe}}\\
&\simeq& 33.6 a_{-5}^{-2}\left( \frac{10^{-8}\cm^{-3}}{n_{\rm CRe}}\right)\left(\frac{0.1c}{v_{\rm CRe}}\right)\yr.\label{eq:tau_CRe}
\ena

The timescale of UV absorption that induces photoemission is given by
\bea
\tau_{\rm UV}&=&\frac{1}{\pi a^{2}\Phi_{\rm UV}}\\
&\simeq& 0.3a_{-5}^{-2}\left(\frac{10^{-17}\s^{-1}}{\zeta({\rm H}_2)}\right)~\yr,\label{eq:tau_UV}
\ena
where $\Phi_{\rm UV}$ is given by Eq. \ref{eq:PHIUV_CR}.

Comparing Equations (\ref{eq:tau_CRe}) and (\ref{eq:tau_UV}) to Equation \ref{eq:tauLar}, one can see that the spin alignment timescale by the Barnett effect is much shorter than the timescale of UV absorption and CR bombardment. Therefore, captured electrons are quickly spin-polarized before the ejection by UV and CR bombardments. As a result, electrons ejected by UV/CR bombardment are spin-polarized.

\subsection{Emission of Spin-Polarized Electrons}
\subsubsection{Photoemission by CR-induced UV Radiation}
 Irradiation of dust grains aligned with the ambient magnetic field by unpolarized UV radiation would eject spin-polarized electrons with spins aligned along the magnetic field \citep{Hoang.2024}. The thermal electrons captured by the grain will be easiest to be ejected due to their loosen bound to the grain atoms. Here we study the emission of SPEs from aligned grains by CR-induced UV radiation. 

Let $Y(a,\nu)$ be the photoelectric yield of a grain of size $a$ induced by a photon of frequency $\nu$. The rate of photoelectron emission of primary electrons from one grain (electrons per second) is
\bea
J_{\rm pe}(a)&&=\int_{\nu_{\rm pet}}^{\infty} Y(a,\nu)\pi a^{2}Q_{\rm abs} \frac{cu_{\nu}}{h\nu} d\nu,\label{eq:Jpe}\\
&&= \pi a^{2}\Phi_{\rm UV}\langle Y(a,\nu)Q_{\rm abs}\rangle \nonumber\\
&&\simeq 10^{-8}a_{-5}^{2}(c_{0}+c_{1}R_{\rm V}+c_{2}R_{\rm V}^{2})\left(\frac{\zeta({\rm H}_2)}{10^{-17}\s^{-1}}\right)\nonumber\\
&&\times \langle Y(a,\nu)Q_{\rm abs}\rangle,
\ena
where $Q_{\rm abs}$ is the absorption efficiency, $\nu_{\rm pet}$ is the frequency threshold required for the photoelectric effect, which is determined by the ionization potential (IP), i.e., $h\nu_{\rm pet}=\IP$, and  $u_{\nu}$ is the specific energy density of the radiation field. Here we take the ionization potential $\IP=W=8$ eV for silicate grains (see \citealt{WeingartnerDraine.2001a}). The energy density spectrum $u_{\nu}$ is the UV radiation induced by CRs with the flux $u_{\rm UV}=\int u_{\nu}d\nu = \Phi_{\rm UV}/c$ given by Equation (\ref{eq:PHIUV_CR}), and $\langle YQ_{\rm abs}\rangle= \int YQ_{\rm abs}cu_{\nu}d\nu/\Phi_{\rm UV}$.

For the CR-inducec UV radiation of $h\nu \sim 11.2-13.6$ eV, we adopt $\langle YQ_{\rm abs}\rangle=0.1$ because $Y\sim 0.1$ for aligned grains of $1\mum> a>0.1\mum$ and $Q_{\rm abs}\approx 1$ (\citealt{Hoang.2015,Hoang.2024}).

\begin{figure}
    \centering
    \includegraphics[width=0.5\textwidth]{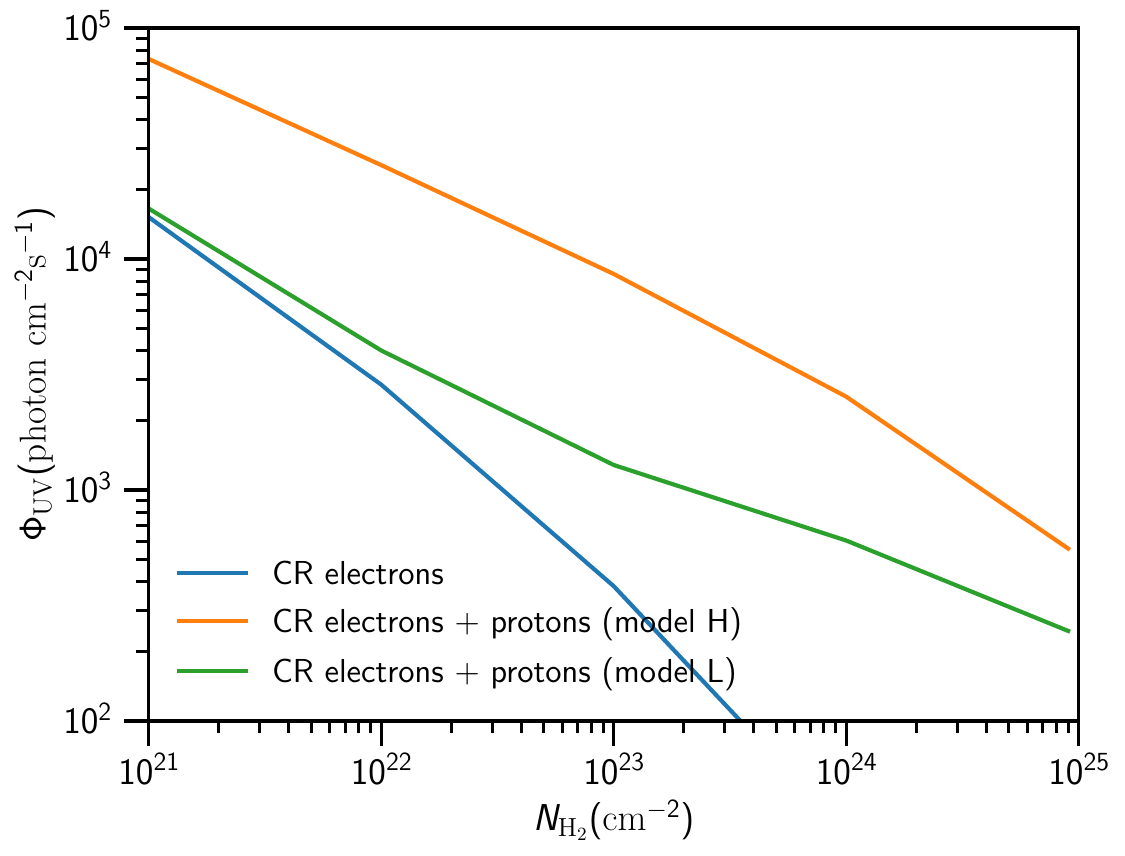}
    \includegraphics[width=0.5\textwidth]{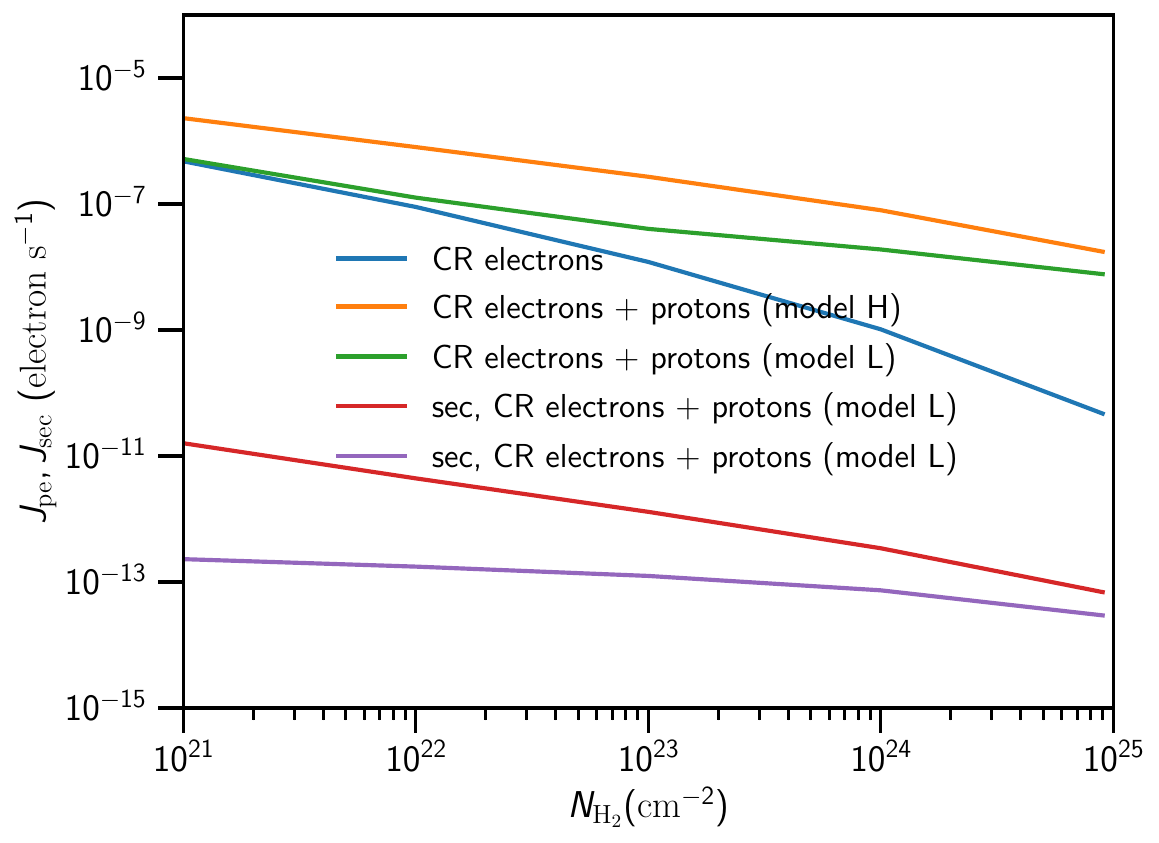}
    \caption{The emission rate of photoelectrons by CR-induced UV radiation and secondary electrons by CR bombardment. Photoemission is dominant over secondary electron emision. The grain size of $a=0.1\mum$ is adopted.}
    \label{fig:jpe}
\end{figure}

Figure \ref{fig:jpe} shows the flux of CR-induced UV radiation (upper panel) and the rate of photoemission (lower panel) as functions of the column density. The effect of CR protons is more important in producing UV radiation and then photoemission than CR electrons.

\subsubsection{Secondary electron emission by CRs}
Irradiation of dust grains by CR electrons and protons can create secondary electrons in the inner shell of atoms. Such energetic electrons can cause the ejection of electrons on the way to the grain surface, producing a higher abundance of electrons, an effect similar to the Auger effect caused by X-rays \citep{Hoang.2015}. The secondary electrons usually have low energy \citep{Arumainayagam.2010}. These low-energy secondary electrons include mostly include unpaired electrons because they have lower ionization potential. 

The rate of secondary electron emission from the grain (electrons per second) by CRs is given by
\bea
J_{\rm sec,k}(a,Z)=\pi a^{2}\int_{E_{\rm int}}^{\infty} 4\pi j_{k}(E)\delta_{e}(E)dE,
\ena
where $k=e, p$, $\delta_{e}$ is the secondary photoelectric yield \citep{Draine.1979}, which can be approximately given by\cite{Hoang:2017hg}
\bea
\delta_{e}(E)=4\delta_{e}^{\rm max}\left(\frac{E}{E_{\rm max}}\right)\frac{1}{(1 + E/E_{\rm max})^{2}},\label{eq:delta_e}
\ena
where $\delta_{e}^{\rm max}\sim 1.5-2$, $E_{\rm max}\sim 0.2-0.4$ keV (cf. \citealt{Ivlev.2015a}). The above equation implies that the secondary emission yield is $\delta_{e}\sim \delta_{\rm max}$ at $E=E_{\rm max}$, and it decreases rapidly for $E\gg E_{\rm max}$ as $\delta_{e}/\delta_{\rm max}\sim 4(E/E_{\rm max})^{-1}\sim 0.012(E_{\rm max}/0.3{\rm keV})(100 {\rm keV}/E)$.

The rate of secondary electron emission by CR electrons and protons is shown in Figure \ref{fig:jpe} (lower panel). The secondary electron emission by CRs is negligible compared to photoemission by CR-induced UV radiation.

\subsubsection{Fraction of spin-polarized photoelectrons and secondary electrons}
Photoelectrons and secondary electrons emitted by CR effects mostly arise from electrons in the outermost electronic shells (valence shell) that are the most loosely bound to the nucleus. Now, we estimate the fraction of SPEs among photoelectrons and secondary electrons.

First, for photoemission of unpaired thermal electrons collected by aligned grains, the fraction of SPEs among photoelectrons is $f_{\rm SPE}=100\%$ because captured electrons are most likely unpaired because they are loosely bound to atoms on the outermost energy shell.

Second, if photoelectrons are emitted from the original dust grain, the fraction of SPEs can be estimated as
\bea
f_{\rm SPE}=\frac{N_{\rm unpair}}{N_{\rm valence}}.\label{eq:fSPE}
\ena

For the silicate core of the typical olivine structure MgFeSiO$_{4}$, the number of valence electrons is $n_{\rm valence}=32$, and the number of unpaired electrons is $n_{\rm unpair}=4$, which yields ratio of unpaired to valence electrons is $f_{\rm SPE}=1/8$ per molecule.\footnote{In the MgFeSiO4 molecule, the ionic bonds between Fe$^{2+}$-O$^{2-}$ and Mg$^{2+}$-O$^{2-}$, and covalent bonds between Si-O. The number of valence electrons include 2 electrons from Mg$^{2+}$, 2 electrons from Fe$^{2+}$, 4 electrons from Si, and 6 electrons in the outermost shell ($2s^{2}2p^{4}$) from each O, which sums up to 32 valence electrons. Note that both Mg and Fe have two electrons in the outermost shell $3s^{2}$and $4s^{2}$ so that they can donate 2 electrons for O to fill the orbital $2p^{4}$, making ionic bonds.} For metallic iron with 8 electrons in the outermost shell ($4s^{2}3d^{6}$) and 4 unpaired electrons in $3d^{6}$, one has $f_{\rm SPE}=4/8=1/2$ per iron atom. Table \ref{tab:material_fSPE} shows the results for several typical silicate compositions.

\begin{table*}
	\centering
	\caption{Fraction of unpaired electrons to valence in popular molecules}
	\begin{tabular}{l|l|l|l}
		Molecules & Valence Electrons, $n_{\rm val}$ & Unpaired Electrons, $n_{\rm unpair}$ & Fraction, $f_{\rm SPE}$\cr
		\hline\cr
		Olivine (MgFeSiO$_4$) & 2(Mg)+ 2(Fe)+ 4(Si)+24 (O4)= 32 & 4 & 1/8 \cr
		Pyroxene (FeSiO$_3$)  & 2(Fe)+ 4(Si) + 18 (O3)=24  & 4 & 1/6 \cr
		Metallic Iron (Fe) & 8 & 4 & 1/2 \cr
		W$\ddot{u}$stite (FeO) & 2(Fe)+ 6 (O)=8 & 4 & 1/2 \cr
		Hematite (Fe$_2$O$_3$)   & 4(Fe2) + 18 (O3)=22 & 10 & 5/11 \cr
		Maghemite (Fe$_3$O$_4$)  & 6 (Fe3)+ 24 (O4)=30 & 14 & 7/15 \cr
		Carbon (C) & 4 & 0 & 0 \cr
		\hline
		\label{tab:material_fSPE}
	\end{tabular}
\end{table*}

For the silicate structure of MgFeSiO$_{4}$ with embedded iron inclusions, let $\phi_{\rm Fe}$ be the volume filling factor of metallic iron. The volume of metallic iron is $V_{\rm Fe}=\phi_{\rm Fe}V_{\rm grain}$ and the volume of silicate is $V_{\rm silicate}= (1-\phi_{\rm Fe})V_{\rm grain}$ where $V_{\rm grain}=4\pi a^{3}/3$.
 
We have the total number of valence and unpaired electrons in the grain of size $a$ are
\bea
N_{\rm valence} &&= n_{\rm val}(Fe)\frac{\phi_{\rm Fe}V_{\rm grain}}{v_{\rm Fe}} + n_{\rm val}(Sil)\frac{(1-\phi_{\rm Fe})V_{\rm grain}}{v_{\rm Sil}},\\
N_{\rm unpair} &&= n_{\rm unpair}(Fe)\frac{\phi_{\rm Fe}V_{\rm grain}}{v_{\rm Fe}} + n_{\rm unpair}(Sil)\frac{(1-\phi_{\rm Fe}V_{\rm grain}}{v_{\rm Sil}},
\ena 
where $v_{\rm Fe}=4\pi a_{\rm Fe}^{3}/3=1.15\times 10^{-23}\cm$ with $a_{\rm Fe}\approx 1.4\AA$ is the Fe atom volume, and $v_{\rm Sil}=3.5\times 10^{-23}\cm$ is the volume of the silicate molecule MgFeSiO$_4$. 

Using Equation \ref{eq:fSPE} one obtain $f_{\rm SPE}=0.125$ for $\phi_{\rm Fe}=0$ (i.e., only silicate) and $f_{\rm SPE}=0.22$ for silicate with embedded iron clusters at the maximum Fe volume filling factor of $\phi_{\rm Fe}=0.3$ \citep{HoangLaz.2016}. Therefore, the fraction of SPEs is  about $10\%$ for silicate grains with embedded irons. 

When the value of $f_{\rm SPE}$ is available in Table \ref{tab:material_fSPE}, one can calculate the total emission rate of SPEs from the aligned grain as
\bea
J_{\rm SPE}=f_{\rm SPE}\left( J_{\rm pe} + J_{\rm sec}\right),\label{eq:Jem_SPE}
\ena
where the second term is subdominant than the first term.

Using the photoemission rate shown in Figure \ref{fig:jpe} and the values of $f_{\rm SPE}$ in Table \ref{tab:material_fSPE}, one can estimate the photoemission rate of SPEs is $J_{\rm SPE}\sim 10^{-10} f_{\rm SPE}\s^{-1}$. The flux of SPEs is then $\Phi_{\rm SPE}^{\rm em}=J_{\rm SPE}/(4\pi a^{2})\sim 0.08f_{\rm SPE}a_{-5}^{-2} \cm^{-2}\s^{-1}$.

\section{Production of Secondary SPEs within Aligned Grains by CRs}\label{sec:SPE_production}
Secondary SPEs produced within an aligned grain by CR bombardment play an important role for chemistry in the icy grain mantle. Here, we estimate the rate of secondary SPEs produced by CR electrons and protons.
\subsection{Range of CR electrons and protons in dust}
Above, we have studied the collection of electrons and emission of photoelectrons and secondary electrons from aligned grains. Because UV photons have attennuation length $l_{a}\sim \lambda/(4\pi \Im(m))$ with $m$ the refractive index of the dust at wavelength $\lambda$ with a typical value of $l_{a}\approx 0.01\mum$ \citep{WeingartnerDraine.2001a}, UV photons can cause photoemission of electrons from the surface layer of $0.01\mum$ of the grain. However, CR electrons/protons have the range (see \citep{Hoang.2015})
\bea
R_{e}&& \simeq 0.01 \hat{\rho}^{-0.85}(E_{e}/1{~\rm keV})^{1.5}\mum,\\
R_{p}&& \simeq 0.01\hat{\rho}^{-1}(E_{p}/1{\rm keV})\mum,\label{eq:Range_ep}
\ena
which can excite electrons deep inside the grain. 

We note that only the secondary electrons that have sufficient energy can travel from the grain core to the surface and be able to escape from the grain. The remaining secondary electrons cannot escape the grain, but they can still trigger chemical reactions of radical and simple molecules to form COMs and amino acids, leading to chiral asymmetry. Here, we first estimate the fraction of SPEs created inside an aligned grains by CR electrons and protons.

\subsection{Flux of secondary SPEs produced by CR electrons and protons}
Upon passing through the dust grains, CR electrons and protons loose their energy through various electronic processes including ionization and excitation (see \citealt{Hoang.2015} for details). The energy loss is given in Appendix \ref{eq:dEdx_el}. 


Let $w$ be the average energy required to create an electron-hole pair within the grain, which has the value of $\sim 10-30\ev$ for silicate. Assuming the effective spherical grain of size $a$ as a slab of thickness $4a/3$, the rate of SPEs produced by an aligned grain within the protostellar core to that the CR traverses the hydrogen column density $N=N_{\H_{2}}$:
\bea
R_{\rm SPE}^{\rm sec,k}(N)= \pi a^{2}f_{\rm SPE}\int_{E_{\rm cut}}^{E_{\rm max}} \frac{4a}{3w}\left.\frac{dE}{dx}\right\vert_{E'} 4\pi j_{k}(E',N) dE',~~~
\label{eq:Rsec}
\ena
where $j_{k}(E',N)$ with $k=e, p$ is the spectrum of CR electrons and protons at the column density $N$ (see Fig. \ref{fig:jCR}).

The flux of SPEs propgating through the grain (in unit of electrons per $\cm^{2}$ per second) can be calculated as
\bea
\Phi_{\rm SPE}^{\rm sec,k}(N)&&=\frac{R_{\rm sec}^{\rm SPE}}{4\pi a^{2}}\nonumber\\
&&=f_{\rm SPE}\frac{4\pi a}{3w}\int_{E_{\rm cut}}^{E_{\rm max}}  4\pi j_{k}(E',N) \left.\frac{dE}{dx}\right\vert_{E'}dE'.~~~~
\label{eq:Phi_SPE_CRe}
\ena

\begin{figure}
	\includegraphics[width=0.5\textwidth]{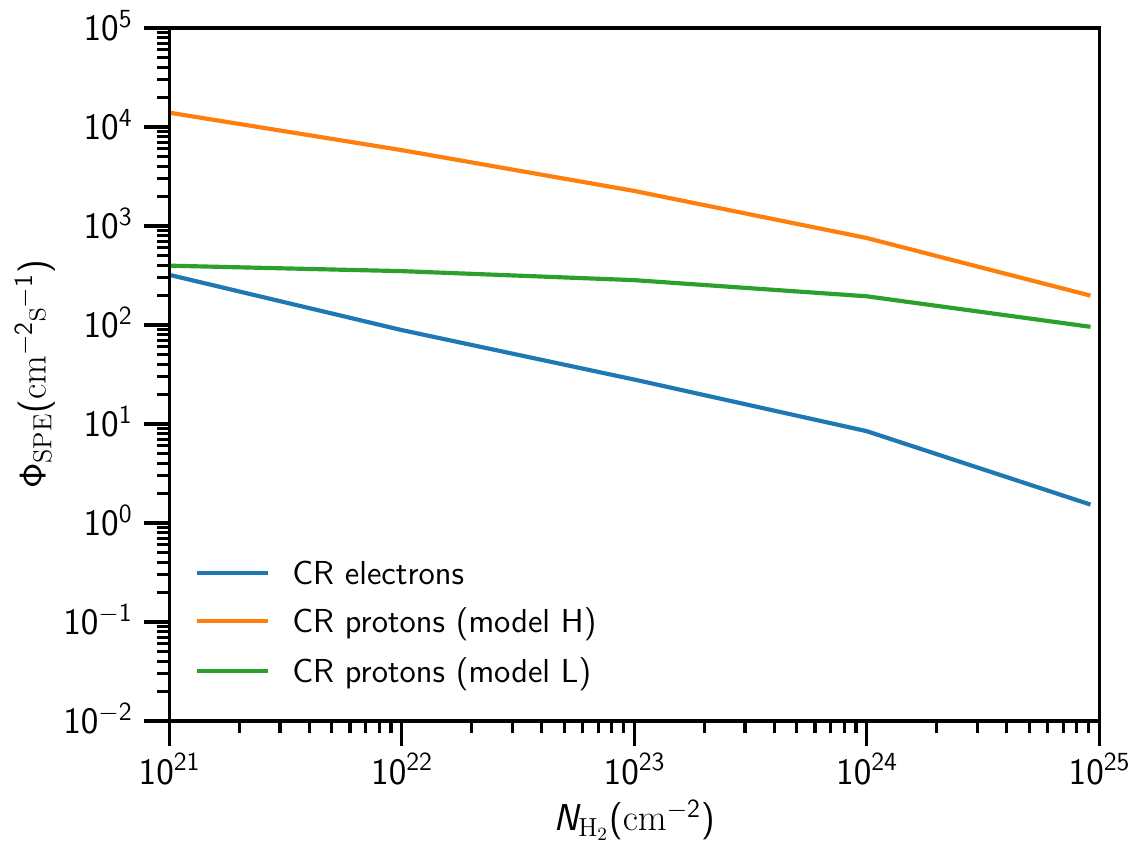}
	\caption{Flux of secondary SPEs emitted by CR electrons and protons bombarding silicate grains, assuming $f_{\rm SPE}=0.1$. CR protons are dominant over CR electrons in secondary electron emission.}
	\label{fig:PHI_sec_SPE}
\end{figure}

Using the attenuated spectra of CR electron and protons calculated at the gas column density $N_{\H_{2}}$, (see Fig. \ref{fig:jCR}) and the energy loss $dE/dx$ in Appendix \ref{apdx:energy_loss}, we calculate the flux of secondary SPEs created within the silicate grain and show the results in Figure \ref{fig:PHI_sec_SPE}. The flux of SPEs by CR protons is dominant over that of CR electrons for both model H and L. For CR electrons, one see that $\Phi_{\rm SPE}^{\rm sec, CRe}\approx 100, 50, 10\cm^{-2}\s^{-1}$ at $N_{\H_{2}}=10^{22}, 10^{23}, 10^{24}\cm^{-2}$, assuming the typical value of $f_{\rm SPE}=0.1$ and the grain size $a=1\mum$. However, for CR protons (model L), one has $\Phi_{\rm SPE}^{\rm sec, CRp}\approx 100\cm^{-2}\s^{-1}$ at $N_{\H_{2}}=10^{22}, 10^{23}, 10^{24}\cm^{-2}$ using the same parameters. 

Compared to UV flux produced by CRs obtained in Figure \ref{fig:jpe} (upper panel), the flux of secondary SPEs (Fig. \ref{fig:PHI_sec_SPE}) is generally lower. However, the reaction cross-section of electrons with molecules is larger than photons \citep{Arumainayagam.2010} (see \cite{Wu.2024} for more discussion). Therefore, secondary electrons and SPEs would be important for chiral chemistry in the ice mantle.

\section{Discussion}
\label{sec:discuss}
\subsection{Effects of CRs on production of low-energy SPEs from Aligned Grains}
In Paper I \citep{Hoang.2024}, we first studied the origins and effect of spin-polarized electrons on chiral symmetry breaking. We suggested that dust grains aligned with the ambient magnetic fields are a ubiquitous source of spin-polarized electrons (SPEs) due to photoelectric effect caused by interstellar UV radiation. We also suggest that magnetically aligned grains could play an important role in chiral symmetry breaking of prebiotic molecules (amino acids and sugars) due to chiral-dependence adsorption of chiral molecules onto aligned grains.

Here, we have studied the effects of CRs on emission of SPEs from aligned dust grains in dense protostellar environments where interstellar UV radiation is significantly reduced. Using the popular CSDA model of CR transport, we calculate the spectra of CRs for different gas column densities and the flux of CR-induced UV radiation. We found that thermal electrons produced by CR ionization that have random spins captured to aligned grains become spin-polarized due to the Barnett effect because the timescale of the spin alignment is shorter than the electron collision time (see Section \ref{sec:SPE_emission}). Using the flux of CR-induced UV radiation, we calculate the rate of photoemission and found that CR-induced UV is dominant in producing photoelectrons from the aligned grains, including spin-polarized electrons (SPEs) (see Figures \ref{fig:jpe}). CR electrons and protons can also create secondary SPEs within the silicate core of the icy grain (see Figure \ref{fig:PHI_sec_SPE}). 

We note that the depolarization of SPEs is negligible for low-energy regime as shown in \cite{Hoang.2024}, so SPEs can maintain their spin orientation during the passage through the grain and in the gas.

\subsection{Role of low-energy electrons in the formation of amino acids in ice mantles}
As shown in this paper, CRs are the dominant source of ionizing molecular gas, $\H_{2}$, producing low-energy electrons in protostellar environments. Moreover, CR bombardment can also produce secondary low-energy electrons, althought its flux is lower.

Experiments have demonstrated that low-energy electrons play a vital role in surface chemistry. For instance, irradiation of low-energy electrons onto the ice mantles analogous to interstellar ice was found to trigger chemical reactions in similar ways as irradiation of UV photons \cite{Boamah.2014,Boyer.2016,Sullivan.2016,Kipfer.2024}. In particular, the experiment by \cite{Esmaili.2018} showed that glycine can be formed by irradiation of low-energy electrons on interstellar analog of CO$_{2}$-CH$_{4}$-NH$_{3}$ ice (see \citealt{Arumainayagam.2019} for a review). 
Recently, \cite{Wu.2024} calculated the flux of secondary electrons by CR protons and suggest that due to their larger cross-section than UV photons, secondary low-energy electrons play a more important role in ice chemistry than UV photons. 



\subsection{Can low-energy spin-polarized electrons induce the chiral asymmetry of amino acids formed in the ice mantle?}
The role of low-energy electrons on ice chemistry is well established. Yet, to date, the effect of spin of low-energy electrons on the chirality of prebiotic molecules formed in interstellar ice analogs is not yet studied. It is well established from experiments that the chiral asymmetry of amino acids could be achieved by irradiation of UVCPL (e.g., \citealt{Modica.2014}). Based on the similarity between UVCPL and SPEs in terms of their induced helicity described by $H\propto \bS.\bv$, we conjecture that SPEs have an effect on the chiral asymmetry of molecules formed in the ice mantle of aligned grains.

We note that secondary SPEs created within the aligned grains by CRs can travel an average distance, defined by {\it electron escape length, $l_{e}$}. The typical value of $l_{e}$ for low-energy electrons of $\sim 10$ eV through silicate grains is $l_{e}\sim 10\AA$ (see \citealt{WeingartnerDraine.2001a} for more details). Therefore, they can reach the outer ice mantle and trigger chemical activities on the ice mantle. Moreover, compared to UV-CPL, which is the leading mechanism for causing chiral symmetry breaking, low-energy SPEs produced by CR particles are more abundant due to the significant attenuation of UV photons. Similarly, the cross-section of electrons is larger photons. Therefore, we predict that low-energy SPEs from CRs dominate the chiral symmetry breaking in icy grain chemistry in dense protostellar environments.

\subsection{A model for origins of chiral asymmetry of amino acids in protostellar environments} 
\begin{figure*}
    \centering
\includegraphics[width=0.9\textwidth]{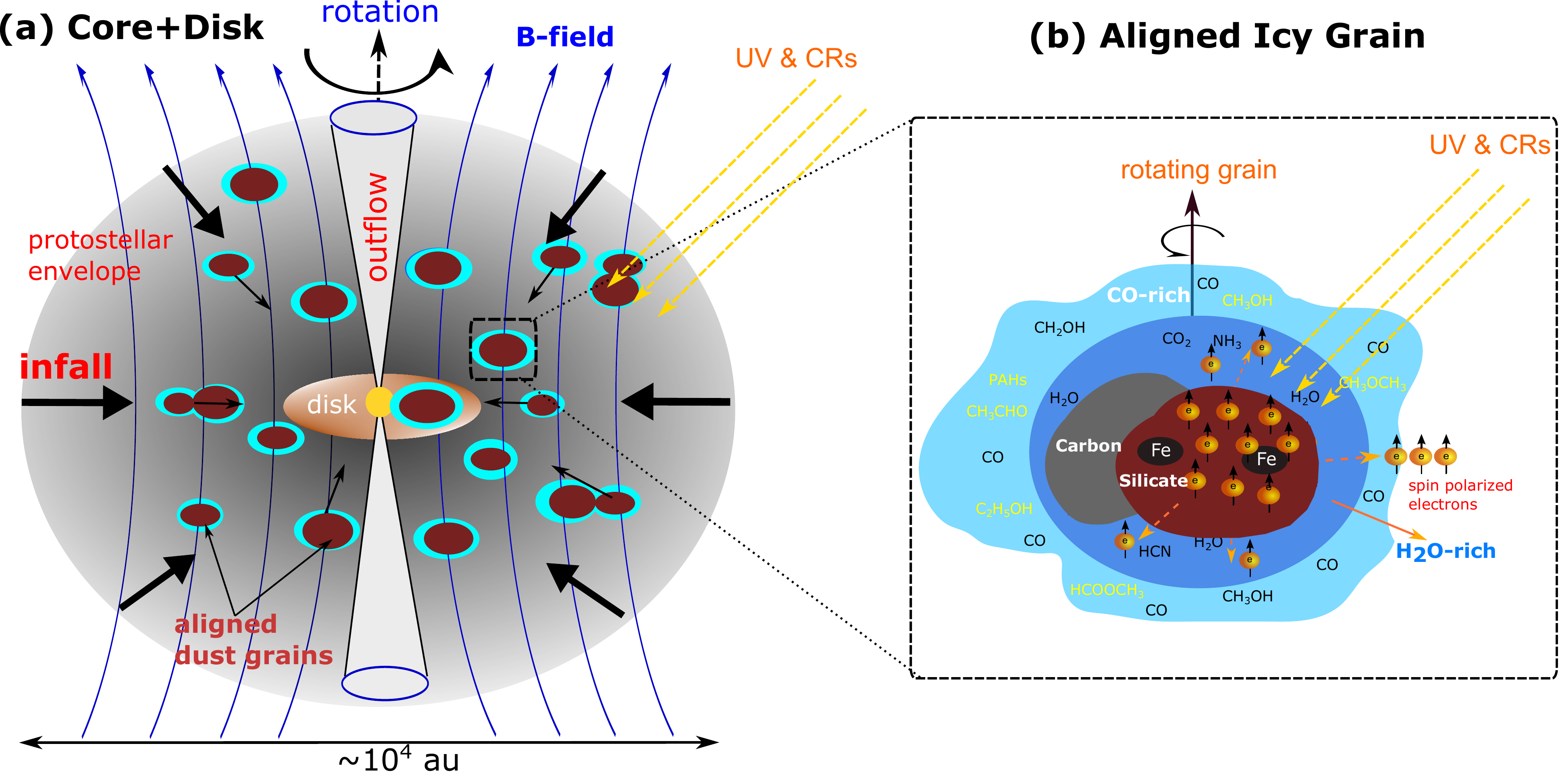}
    \caption{(a)Illustration of a protostellar system including envelope and the disk. Icy dust grains are aligned with magnetic fields. (b) Zoom-in a single icy grain. SPEs produced by CRs and CR-induced UV radiation impact chemical reactions and the formation of amino acids within the ice mantles. Irradiation of SPEs on molecules in the ice mantle can facilitate the formation of COMs and induce chiral asymmetry.}
    \label{fig:protostar_icegrain_SPE}
\end{figure*}

Icy grain mantles play an essential role in the formation of COMs, amino acids (see review by \citealt{Herbst:2009go}), and sugars \citep{Meinert.2016}. Here, following Paper I, we suggest that icy magnetic grains aligned with the ambient magnetic field in protostellar systems could be a key agent for the symmetry breaking of chiral prebiotic molecules. The scenario is illustrated in Figure \ref{fig:protostar_icegrain_SPE}.

First, aligned grains produce low-energy SPEs due to the photoelectric effect caused by interstellar UV photons \citep{Hoang.2024} and CR-induced UV radiation. Secondary SPEs created inside the grain by CR bombardment would travel to the ice mantle and have important effects on the chemical synthesis within the ice mantle (see Figure \ref{fig:icemantle_SPE}), as shown by experiments with electron irradiation on interstellar ice analogs. When accounting for the electron spin, SPEs differently interact with chiral molecules on the aligned grain ice mantle, causing the chiral asymmetry due to the dependence of reaction rates of chiral molecules on electron spins \citep{Rosenberg.2008,Rosenberg.2019,Ozturk.2022}. 

Second, aligned icy grains containing aligned electron spins could act as the magnetic surface that induces the chiral-selective adsorption of chiral molecules in the gas phase due to exchange interaction \citep{Banerjee-Ghosh.2018,Ozturk.2023b}, leading to the accumulation of a preferred enantiomer of chiral molecules than the other and chiral asymmetry. Finally, aligned grains can act as a chiral agent which helps form more complex molecules of similar chirality and increase chiral asymmetry thanks to magnetic dipole-dipole interaction \citep{Hoang.2024}. Finally, the resulting SPEs induced by CR-induced UV radiation will escape into the gas and interact with nearby icy grains before they are captured by hydrogen molecules, as shown in \citep{Hoang.2024}. This flow of SPEs can cause the chiral asymmetry on icy grains in planet-forming disks and cause the initial enantiomer excess.

In our proposed scenario, amino acids formed intrinsically chiral asymmetry in aligned icy grains. Ice grains with amino acids become building blocks of planets, comets, and asteroids. Some amino acids from comets/asteroids may be delivered to Earth and act as origins of life. Amino acids in the ice mantle are racemic, but those adsorb/crystalize onto the grain core are chiral asymmetric due to the spin exchange interaction \citep{Rosenberg.2008,Tassinari.2019}. The advantage of this model is that chiral asymmetry is protected from further destruction due to the shield from energetic radiation/particles. In particular, our model can explain the leading explanation for the extraterrestrial origin of amino acids in carbonaceous meteorites that were formed from the ISM as demonstrated by isotopic analysis of deuterium \citep{Engel.1997,Pizzarello.2005,Glavin.2020,Glavin.2020wwc}.

Note that our proposed model is different from the model by \cite{Greenberg.1996}. In Greenberg's model, amino acids are formed in the ice mantle of grains in molecular clouds. Subsequently, the clouds pass through the UV circularly polarized light from neutron stars. However, to date, no measurements of UVCPL from neutron stars are reported (see \citep{Bailey.2001} for more discussion). Our proposed model here shares the similarity with Greenberg's model by assuming amino acids formed in the ice mantle, but it relies on the in-situ SPE from the grain core by CR bombardment to cause the symmetry breaking of amino acids.

Finally, amino acids and sugars in objects such as planetesimals, comets, and asteroids may become asymmetric due to the effect of SPEs emitted from aligned grains with large iron clusters \citep{Hoang.2022} because CRs can penetrate tens of meters within the object and create secondary SPEs. This has implications for the origin of life in the universe.

\subsection{How to observe chiral molecules and chiral asymmetry in astrophysical environments?}
Chiral organic molecules can be detected through the vibrational transitions in the C-C and C-H bonds, which occurs in mid-IR. A more convenient technique is based on the rotational transitions of molecules in radio (rotational spectroscopy). The first chiral molecule, prolyline oxide, is detected in a shell around the massive protostar in the Sgr B2 star-forming region by \cite{McGuire.2016} using rotational spectroscopy observed at 11 GHz by GBT, but the enantiomer excess of this molecule is unknown.

The unique method to detect chirality (i.e., handedness) of chiral molecules is using the circular polarization of light. Chiral molecules absorb or scatter dominantly the left-handed/right-handed CP light due to the interaction of the electric field of light with the chiral molecular \citep{Sparks.2009,Sparks.2009md9,Patty.2021,Gleiser.2022}. For example, linearly polarized light will be rotated by the absorption or scattering of chiral molecules. Therefore, by observing the CP of scattered light, one can infer the chirality. \cite{Sparks.2009} detected CP of light scattered by photosynthetic microbes. \cite{Lankhaar.2022d3} calculated the CP signal expected from a simple amino acid detected toward the GC and found the signal is rather small. It is thus very challenging to detect such a small signal due to a small enantiomer excess of amino acids. The most viable method detecting chiral assymmetry of prebiotic molecules is through in-situ measurements of dust in the Zodiacal cloud and from active commets/asteroids and direct analysis of meteorite. 


\subsection{Effects of electron spin alignment and grain helicity on chiral symmetry breaking caused by SPEs}
	So far, we have qualitatively considered the potential role of SPEs from aligned grains on the chiral asymmetry of amino acids. Here, we discuss additional effects that must be considered for accurate determination of SPE polarization and its effect on chiral symmetry breaking. These include the net alignment of electron spins within an aligned grain and the helicity of dust grains. The latter governs the alignment of the grain angular momentum with the ambient magnetic field \citep{LazHoang.2007}.	
	\subsubsection{Alignment of electron spins by the Barnett field}\label{apdx:spin_align}
	In the equilibrium state, the net alignment direction of electron spins within a rotating dust grain is directed along the applied Barnett field (see Eq. \ref{eq:muBar}). However, thermal fluctuations due to the grain temperature tend to cause fluctuations of individual electron spins with respect to $\bB_{\rm Barnett}$ and strongly influence the alignment degree of electron spins. Here we quantify the net alignment of electron spins with the Barnett field. 
	
	In an ambient magnetic field $\bB_{\rm Barnett}$, the electron magnetic moment $\bmu$ posses a magnetic energy
	\bea
	E=-\bmu.\bB_{\rm Barnett} =\mu B_{\rm Barnett}\cos\theta,\label{eq:Eel_Bar}
	\ena
	where $\theta$ is the angle between the electron spin and $\bB_{\rm Barnett}$.
	
	For the fast Larmor precession of the electron magnetic moment around the Barnett field, the probability of finding the spin within the angle $\theta,\theta+d\theta$ is given by the Boltzmann distribution:
	\bea
	P(\theta) = Z \sin\theta \exp\left(\frac{-E}{kT_{d}}\right)=Z \sin\theta\exp\left(-x\cos\theta\right).\label{eq:Ptheta}
	\ena
	where $T_{d}$ is the dust grain temperature, $Z$ is the normalization constant such as $\int_{0}^{\pi}P(\theta)d\theta=1$, and
	\bea
	x=\frac{\mu B_{\rm Barnett}}{kT_{d}}.
	\ena

	The net alignment degree of electron spins with the Barnett field can be calculated using the distribution function $P(\theta)$:
	\bea
	R_{\rm spin} = \frac{1}{2}\langle 3\cos^{2}\theta-1\rangle.\label{eq:Rspin_align}
	\ena
	where 
	\bea
	\langle \cos^{2}\theta\rangle = \int_{0}^{\pi} \cos^{2}\theta P(\theta)d\theta.\label{eq:cos2theta}
	\ena

	%
	
\begin{figure}
	\includegraphics[width=0.5\textwidth]{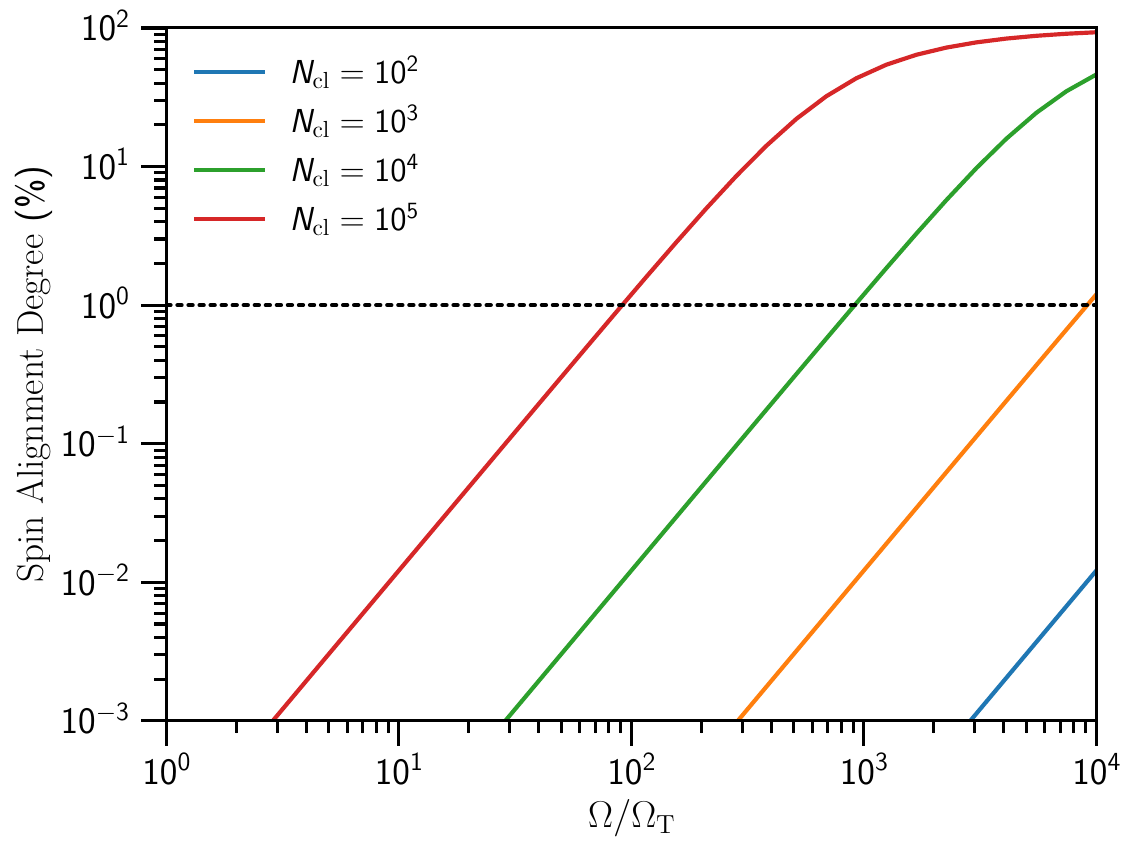}
	\caption{The alignment degree of electron spins with the magnetic fields for grains with embedded iron clusters versus the grain rotation rate for a grain size of $a=0.1\mum$. Different sizes of iron clusters are considered. The combination of suprathermal rotation and large clusters can induce the alignment degree above $1\%$ (dotted back line).}
	\label{fig:Ralign}
\end{figure}

	We calculate the alignment degree $R_{\rm spin}$ of iron clusters in which electron spins are aligned efficiently via exchange interaction. The alignment of electron spins with the Barnett field is just the alignment of the iron cluster's magnetic moments with the Barnett field, and one can take $\mu$ to be the magnetic moment of a cluster $\mu=\mu_{\rm cl}=N_{cl}p\mu_{B}$. Figure \ref{fig:Ralign} shows the alignment degree, for the different values of the grain angular velocity relative to the thermal velocity ($\Omega/\Omega_{T}$), and different sizes of iron clusters ($N_{\rm cl}$), assuming the upper limit for iron inclusions of $\phi_{sp}=0.3$ (\citealt{HoangLaz.2016}). The alignment degree depends increases with the grain rotation rate and the iron cluster size. For the largest cluster of $N_{cl}=10^{5}$, the spin alignment degree is greater than $10\%$ if grains rotate suprathermally with $\Omega>200\Omega_{T}$. Smaller clusters require faster rotation to achieve the similar spin alignment degree. It is noted that paramagnetic grains have negligible spin alignment due to the lower magnetic moment unless grains are rotating extremely fast of $\Omega>10^{4}\Omega_{T}$. However, extremely fast rotating grains would be disrupted into smaller fragments by centrifugal stress \citep{Hoang.2019nas}.

\subsubsection{Helicity of interstellar dust grains}
Above, we have discussed for a particular case of grain shape with right helicity so that electron spins are aligned along the alignment axis determined by the ambient magnetic field. In realistic conditions, grains would have both right and left helicity, as first identified by \cite{LazHoang.2007}. Among 6 shapes considered in \citep{LazHoang.2007}, two shapes (1 and 6) have right helicity, and four shapes (2, 3, 4-5) have left helicity. From a detailed study in \cite{Herranen.2019} for 15 different shapes, we show that the majority of shapes has left helicity. Among 18 porous shapes considered in \cite{Shen.2008ApJ}, we found 11 shapes have left helicity and 7 shapes with right helicity (Hoang et al. in preparation). Therefore, there seems to have an excess in the left-helicity of dust grains. Yet, a comprehensive study for a large number of shapes should be carried out to establish whether the left-helicity is dominant and what physical mechanisms would cause such a symmetry breaking of grain shapes.

\section{Summary}\label{sec:summary}
We studied the effects of CRs and CR-induced radiation on the production of SPEs from aligned grains in dense protostellar environments and discuss their implications for chiral symmetry breaking. Our main findings are summarized as follows:
\begin{enumerate}
\item Assuming the most likely composite dust model for the dense regions consisting of the Astrodust core and an ice mantle, we find that dust grains are efficiently aligned with the magnetic fields by magnetically enhanced radiative torque (MRAT) mechanism in dense protostellar environments. Electron spins are aligned within such magnetically aligned grains with the ambient magnetic field.

\item We find that thermal electrons having random spins captured by aligned grains will become spin-polarized due to the Barnett effect, converting randomly electron spins to aligned spins.

\item Using the continuous slowing down approximation model for modeling the CR transport in dense regions, we calculate the CR-induced UV radiation by H$_2$ fluoresence. We find that local UV radiation induced by induces the photoemission of spin-polarized electrons from aligned dust grains, which are the important source of SPEs in dense environments.

\item CRs can also create secondary SPEs within the grain core, which may trigger chemical reactions and help form complex molecules and amino acids in the icy mantles of the aligned grain.

\item Due to the physical similarity between UV-CPL and SPEs in terms of spins and helicity, we suggest that SPEs produced by CRs could play an important role in surface chemistry in ice mantles of grains and might produce symmetry breaking for resulting amino acids in star- and planet-forming regions.

\item We suggest a novel model for explaining the origins of amino acids detected in meteorite, comets, and asteroids, which is based the effect of SPEs induced by CRs on aligned grains and ice mantles in protostellar environments. Our proposed mechanism has advantage over UVCPL because CRs can penetrate deep into the planet-forming regions. However, the exact level of chiral excess depends on the alignment degree of SPEs in aligned grains and the helicity of dust grains, which requires further studies.

\end{enumerate}

During the review process, results from analysis of the sample return from Bennu primitive asteroid was published in \cite{Glavin.2025}, which shows the almost equal amount of L- and D-amino acids. This unexpected discovery challenges the leading hypothesis for extraterrestrial origin of L-amino acids on the Earth, and the true source of L-amino acids in meteorites remains a mystery. Our results here show that, although SPEs might cause the chiral asymmetry of amino acids/sugars in protostellar environments, their exact efficiency is still uncertain due to the uncertainty in the size of iron clusters embedded inside dust grains and the helicity of grain shapes.

\acknowledgments
We thank the anonymous referee for constructive comments and Phan Vo Hong Minh for discussions on cosmic ray transport in molecular clouds. This work is supported by the research planning for exploring cosmic life phenomena (LiCE) project (No. 2024E84100) funded by Korea Astronomy and Space Science (KASI). This work was partly supported by a grant from the Simons Foundation to IFIRSE, ICISE (916424, N.H.). We would like to thank the ICISE staff for their enthusiastic support.


\bibliography{ms.bbl}

\begin{thebibliography}{133}
\expandafter\ifx\csname natexlab\endcsname\relax\def\natexlab#1{#1}\fi

\bibitem[{Altwegg {et~al.}(2016)Altwegg, Balsiger, Bar-Nun, Berthelier, Bieler,
  Bochsler, Briois, Calmonte, Combi, Cottin, Keyser, Dhooghe, Fiethe, Fuselier,
  Gasc, Gombosi, Hansen, Haessig, Jäckel, Kopp, Korth, Roy, Mall, Marty,
  Mousis, Owen, Rème, Rubin, Sémon, Tzou, Waite, \& Wurz}]{Altwegg.2016}
Altwegg, K., Balsiger, H., Bar-Nun, A., {et~al.} 2016, Science Advances, 2,
  e1600285

\bibitem[{Arumainayagam {et~al.}(2010)Arumainayagam, Lee, Nelson, Haines, \&
  Gunawardane}]{Arumainayagam.2010}
Arumainayagam, C.~R., Lee, H.-L., Nelson, R.~B., Haines, D.~R., \& Gunawardane,
  R.~P. 2010, Surface Science Reports, 65, 1

\bibitem[{Arumainayagam {et~al.}(2019)Arumainayagam, Garrod, Boyer, Hay, Bao,
  Campbell, Wang, Nowak, Arumainayagam, \& Hodge}]{Arumainayagam.2019}
Arumainayagam, C.~R., Garrod, R.~T., Boyer, M.~C., {et~al.} 2019, Chemical
  Society Reviews, 48, 2293

\bibitem[{Bailey(2001)}]{Bailey.2001}
Bailey, J. 2001, Origins of life and evolution of the biosphere, 31, 167

\bibitem[{Bailey {et~al.}(1998)Bailey, Chrysostomou, Hough, Gledhill, McCall,
  Clark, Menard, \& Tamura}]{Bailey.1998}
Bailey, J., Chrysostomou, A., Hough, J.~H., {et~al.} 1998, Science, 281, 672

\bibitem[{Banerjee-Ghosh {et~al.}(2018)Banerjee-Ghosh, Dor, Tassinari, Capua,
  Yochelis, Capua, Yang, Parkin, Sarkar, Kronik, Baczewski, Naaman, \&
  Paltiel}]{Banerjee-Ghosh.2018}
Banerjee-Ghosh, K., Dor, O.~B., Tassinari, F., {et~al.} 2018, Science, 360,
  1331

\bibitem[{Barnett(1915)}]{Barnett.1915}
Barnett, S.~J. 1915, Physical Review, 6, 239

\bibitem[{Bernstein {et~al.}(2002)Bernstein, Dworkin, Sandford, Cooper, \&
  Allamandola}]{Bernstein.2002}
Bernstein, M.~P., Dworkin, J.~P., Sandford, S.~A., Cooper, G.~W., \&
  Allamandola, L.~J. 2002, Nature, 416, 401

\bibitem[{{Bialy} {et~al.}(2022){Bialy}, {Belli}, \& {Padovani}}]{bialy2022}
{Bialy}, S., {Belli}, S., \& {Padovani}, M. 2022, \aap, 658, L13

\bibitem[{Blackmond(2004)}]{Blackmond.2004}
Blackmond, D.~G. 2004, ChemInform, 35, no

\bibitem[{Boamah {et~al.}(2014)Boamah, Sullivan, Shulenberger, Soe, Jacob,
  Yhee, Atkinson, Boyer, Haines, \& Arumainayagam}]{Boamah.2014}
Boamah, M.~D., Sullivan, K.~K., Shulenberger, K.~E., {et~al.} 2014, Faraday
  Discussions, 168, 249

\bibitem[{Bonner(1991)}]{Bonner.1991}
Bonner, W.~A. 1991, Origins of Life and Evolution of the Biosphere, 21, 59

\bibitem[{Boyer {et~al.}(2016)Boyer, Rivas, Tran, Verish, \&
  Arumainayagam}]{Boyer.2016}
Boyer, M.~C., Rivas, N., Tran, A.~A., Verish, C.~A., \& Arumainayagam, C.~R.
  2016, Surface Science, 652, 26

\bibitem[{Caro {et~al.}(2002)Caro, Meierhenrich, Schutte, Barbier, Segovia,
  Rosenbauer, Thiemann, Brack, \& Greenberg}]{Caro.2002}
Caro, G. M.~M., Meierhenrich, U.~J., Schutte, W.~A., {et~al.} 2002, Nature,
  416, 403

\bibitem[{Caselli \& Ceccarelli(2012)}]{Caselli:2012fq}
Caselli, P., \& Ceccarelli, C. 2012, The \aa Review, 20, 151

\bibitem[{{Caselli} {et~al.}(1998){Caselli}, {Walmsley}, {Terzieva}, \&
  {Herbst}}]{caselli1998}
{Caselli}, P., {Walmsley}, C.~M., {Terzieva}, R., \& {Herbst}, E. 1998, \apj,
  499, 234

\bibitem[{{Cristofari}(2021)}]{cristofari2021b}
{Cristofari}, P. 2021, Universe, 7, 324

\bibitem[{Cronin \& Pizzarello(1997)}]{Cronin.1997}
Cronin, J., \& Pizzarello, S. 1997, Science, 81, 73

\bibitem[{Draine \& Salpeter(1979)}]{Draine.1979}
Draine, B.~T., \& Salpeter, E.~E. 1979, \apj, 231, 77

\bibitem[{Elsila {et~al.}(2007)Elsila, Dworkin, Bernstein, Martin, \&
  Sandford}]{Elsila.2007}
Elsila, J.~E., Dworkin, J.~P., Bernstein, M.~P., Martin, M.~P., \& Sandford,
  S.~A. 2007, The Astrophysical Journal, 660, 911

\bibitem[{{Engel} \& {Macko}(1997)}]{Engel.1997}
{Engel}, M.~H., \& {Macko}, S.~A. 1997, \nat, 389, 265

\bibitem[{Esmaili {et~al.}(2018)Esmaili, Bass, Cloutier, Sanche, \&
  Huels}]{Esmaili.2018}
Esmaili, S., Bass, A.~D., Cloutier, P., Sanche, L., \& Huels, M.~A. 2018, The
  Journal of Chemical Physics, 148, 164702

\bibitem[{Fano(1963)}]{1963ARNPS..13....1F}
Fano, U. 1963, Annual Review of Nuclear and Particle Sciences, 13, 1

\bibitem[{Flores {et~al.}(1977)Flores, Bonner, \& Massey}]{Flores.1977}
Flores, J.~J., Bonner, W.~A., \& Massey, G.~A. 1977, Journal of the American
  Chemical Society, 99, 3622

\bibitem[{{Gabici} {et~al.}(2019){Gabici}, {Evoli}, {Gaggero}, {Lipari},
  {Mertsch}, {Orlando}, {Strong}, \& {Vittino}}]{gabici2019}
{Gabici}, S., {Evoli}, C., {Gaggero}, D., {et~al.} 2019, International Journal
  of Modern Physics D, 28, 1930022

\bibitem[{{Gabici} \& {Montmerle}(2015)}]{gabici2015}
{Gabici}, S., \& {Montmerle}, T. 2015, in International Cosmic Ray Conference,
  Vol.~34, 34th International Cosmic Ray Conference (ICRC2015), 29

\bibitem[{{Gaches} \& {Offner}(2018)}]{gaches2018}
{Gaches}, B. A.~L., \& {Offner}, S. S.~R. 2018, \apj, 861, 87

\bibitem[{Galametz {et~al.}(2019)Galametz, Maury, Valdivia, Testi, Belloche, \&
  Andr{\'e}}]{Galametz:2019fj}
Galametz, M., Maury, A.~J., Valdivia, V., {et~al.} 2019, \aa, 632, A5

\bibitem[{Giang \& Hoang(2024)}]{GiangHoang.2024}
Giang, N.~C., \& Hoang, T. 2024, Monthly Notices of the Royal Astronomical
  Society, 530, 984

\bibitem[{Giang {et~al.}(2023)Giang, Hoang, Kim, \& Tram}]{GiangHoang.2023}
Giang, N.~C., Hoang, T., Kim, J.-G., \& Tram, L.~N. 2023, Monthly Notices of
  the Royal Astronomical Society, 520, 3788

\bibitem[{Glavin {et~al.}(2020{\natexlab{a}})Glavin, Burton, Elsila, Aponte, \&
  Dworkin}]{Glavin.2020}
Glavin, D.~P., Burton, A.~S., Elsila, J.~E., Aponte, J.~C., \& Dworkin, J.~P.
  2020{\natexlab{a}}, Chemical Reviews, 120, 4660

\bibitem[{Glavin {et~al.}(2012)Glavin, Elsila, Burton, CALLAHAN, DWORKIN,
  HILTS, \& HERD}]{Glavin.2012}
Glavin, D.~P., Elsila, J.~E., Burton, A.~S., {et~al.} 2012, Meteoritics \&
  Planetary Science, 47, 1347

\bibitem[{Glavin {et~al.}(2020{\natexlab{b}})Glavin, McLain, Dworkin, Parker,
  Elsila, Aponte, Simkus, Pozarycki, Graham, Nittler, \&
  Alexander}]{Glavin.2020wwc}
Glavin, D.~P., McLain, H.~L., Dworkin, J.~P., {et~al.} 2020{\natexlab{b}},
  Meteoritics \& Planetary Science, 55, 1979

\bibitem[{Glavin {et~al.}(2025)Glavin, Dworkin, Alexander, Aponte, Baczynski,
  Barnes, Bechtel, Berger, Burton, Caselli, Chung, Clemett, Cody, Dominguez,
  Elsila, Farnsworth, Foustoukos, Freeman, Furukawa, Gainsforth, Graham,
  Grassi, Giuliano, Hamilton, Haenecour, Heck, Hofmann, House, Huang, Kaplan,
  Keller, Kim, Koga, Liss, McLain, Marcus, Matney, McCoy, McIntosh, Mojarro,
  Naraoka, Nguyen, Nuevo, Nuth, Oba, Parker, Peretyazhko, Sandford, Santos,
  Schmitt-Kopplin, Seguin, Simkus, Shahid, Takano, Thomas-Keprta, Tripathi,
  Weiss, Zheng, Lunning, Righter, Connolly, \& Lauretta}]{Glavin.2025}
Glavin, D.~P., Dworkin, J.~P., Alexander, C. M.~O., {et~al.} 2025, Nature
  Astronomy, 1

\bibitem[{Gleiser(2022)}]{Gleiser.2022}
Gleiser, M. 2022, Origins of Life and Evolution of Biospheres, 52, 93

\bibitem[{Greenberg(1996)}]{Greenberg.1996}
Greenberg, J.~M. 1996, AIP Conference Proceedings, 185

\bibitem[{Greenberg \& Li(1998)}]{Greenberg.1998}
Greenberg, J.~M., \& Li, A. 1998, \aa, 332, 374

\bibitem[{Hegstrom {et~al.}(1980)Hegstrom, Rein, \& Sandars}]{Hegstrom.1980}
Hegstrom, R.~A., Rein, D.~W., \& Sandars, P. G.~H. 1980, The Journal of
  Chemical Physics, 73, 2329

\bibitem[{Hensley \& Draine(2021)}]{Hensley.2021}
Hensley, B.~S., \& Draine, B.~T. 2021, The Astrophysical Journal, 906, 0

\bibitem[{Herbst \& van Dishoeck(2009)}]{Herbst:2009go}
Herbst, E., \& van Dishoeck, E.~F. 2009, \araa, 47, 427

\bibitem[{Herranen {et~al.}(2019)Herranen, Lazarian, \& Hoang}]{Herranen.2019}
Herranen, J., Lazarian, A., \& Hoang, T. 2019, \apj, 878, 96

\bibitem[{Hoang(2022)}]{Hoang.2022}
Hoang, T. 2022, \apj, 928, 102

\bibitem[{{Hoang}(2024)}]{Hoang.2024}
{Hoang}, T. 2024, \apj, 976, 26

\bibitem[{Hoang \& Lazarian(2008)}]{HoangLaz.2008}
Hoang, T., \& Lazarian, A. 2008, \mnras, 388, 117

\bibitem[{Hoang \& Lazarian(2016{\natexlab{a}})}]{HoangLaz.2016}
Hoang, T., \& Lazarian, A. 2016{\natexlab{a}}, \apj, 831, 159

\bibitem[{Hoang \& Lazarian(2016{\natexlab{b}})}]{HoangLaz.2016b}
Hoang, T., \& Lazarian, A. 2016{\natexlab{b}}, \apj, 821, 91

\bibitem[{Hoang {et~al.}(2015)Hoang, Lazarian, \& Schlickeiser}]{Hoang.2015}
Hoang, T., Lazarian, A., \& Schlickeiser, R. 2015, \apj, 806, 255

\bibitem[{Hoang \& Loeb(2017)}]{Hoang:2017hg}
Hoang, T., \& Loeb, A. 2017, \apj, 848, 0

\bibitem[{Hoang {et~al.}(2019)Hoang, Tram, Lee, \& Ahn}]{Hoang.2019nas}
Hoang, T., Tram, L.~N., Lee, H., \& Ahn, S.-H. 2019, \natas, 3, 766

\bibitem[{Hoang {et~al.}(2021)Hoang, Tram, Lee, Diep, \& Ngoc}]{Hoang.2021}
Hoang, T., Tram, L.~N., Lee, H., Diep, P.~N., \& Ngoc, N.~B. 2021, \apj, 908,
  218

\bibitem[{{Hoang} {et~al.}(2022){Hoang}, {Tram}, {Minh Phan}, {Giang},
  {Phuong}, \& {Dieu}}]{Hoangetal.2022}
{Hoang}, T., {Tram}, L.~N., {Minh Phan}, V.~H., {et~al.} 2022, \aj, 164, 248

\bibitem[{Hoang \& Truong(2024)}]{HoangBao.2024}
Hoang, T., \& Truong, B. 2024, The Astrophysical Journal, 965, 183

\bibitem[{{Indriolo} \& {McCall}(2012)}]{indriolo2012}
{Indriolo}, N., \& {McCall}, B.~J. 2012, \apj, 745, 91

\bibitem[{Ioppolo {et~al.}(2021)Ioppolo, Fedoseev, Chuang, Cuppen, Clements,
  Jin, Garrod, Qasim, Kofman, Dishoeck, \& Linnartz}]{Ioppolo.2021}
Ioppolo, S., Fedoseev, G., Chuang, K.-J., {et~al.} 2021, Nature Astronomy, 5,
  197

\bibitem[{Ivlev {et~al.}(2015)Ivlev, Padovani, Galli, \& Caselli}]{Ivlev.2015a}
Ivlev, A.~V., Padovani, M., Galli, D., \& Caselli, P. 2015, The Astrophysical
  Journal, 812, 135

\bibitem[{Jenkins(2009)}]{Jenkins.2009}
Jenkins, E.~B. 2009, \apj, 700, 1299

\bibitem[{Kipfer {et~al.}(2024)Kipfer, Galli, Riedo, Tulej, Wurz, \&
  Ligterink}]{Kipfer.2024}
Kipfer, K.~A., Galli, A., Riedo, A., {et~al.} 2024, Icarus, 410, 115742

\bibitem[{Kisker {et~al.}(1982)Kisker, Gudat, \& Schröder}]{Kisker.1982}
Kisker, E., Gudat, W., \& Schröder, K. 1982, Solid State Communications, 44,
  591

\bibitem[{Kvenvolden {et~al.}(1970)Kvenvolden, Lawless, Pering, PETERSON,
  FLORES, PONNAMPERUMA, KAPLAN, \& MOORE}]{Kvenvolden.1970}
Kvenvolden, K., Lawless, J., Pering, K., {et~al.} 1970, Nature, 228, 923

\bibitem[{Kwon {et~al.}(2016)Kwon, Tamura, Hough, Nagata, Kusakabe, \&
  Saito}]{Kwon.2016}
Kwon, J., Tamura, M., Hough, J.~H., {et~al.} 2016, The Astrophysical Journal,
  824, 95

\bibitem[{Kwon {et~al.}(2018)Kwon, Nakagawa, Tamura, Hough, Kandori, Choi,
  Kang, Cho, Nakajima, \& Nagata}]{Kwon.2018}
Kwon, J., Nakagawa, T., Tamura, M., {et~al.} 2018, The Astronomical Journal,
  156, 0

\bibitem[{Kwon {et~al.}(2019)Kwon, Stephens, Tobin, Looney, Li, van~der Tak, \&
  Crutcher}]{Kwon.2019}
Kwon, W., Stephens, I.~W., Tobin, J.~J., {et~al.} 2019, \apj, 879, 25

\bibitem[{Lankhaar(2022)}]{Lankhaar.2022d3}
Lankhaar, B. 2022, Astronomy \& Astrophysics, 666, A126

\bibitem[{Lazarian(2007)}]{Lazarian.2007}
Lazarian, A. 2007, J. Quant. Spectrosc. Rad. Trans., 106, 225

\bibitem[{{Lazarian} {et~al.}(2015){Lazarian}, {Andersson}, \&
  {Hoang}}]{LAH.2015}
{Lazarian}, A., {Andersson}, B.-G., \& {Hoang}, T. 2015, in Polarimetry of
  stars and planetary systems, ed. L.~{Kolokolova}, J.~{Hough}, \& A.-C.
  {Levasseur-Regourd} ((New York: Cambridge Univ. Press)), 81

\bibitem[{Lazarian \& Hoang(2007)}]{LazHoang.2007}
Lazarian, A., \& Hoang, T. 2007, \mnras, 378, 910

\bibitem[{Lazarian \& Hoang(2008)}]{LazHoang.2008}
Lazarian, A., \& Hoang, T. 2008, \apj, 676, L25

\bibitem[{Lazarian \& Hoang(2019)}]{LazHoang.2019}
Lazarian, A., \& Hoang, T. 2019, \apj, 883, 122

\bibitem[{Lee \& Yang(1956)}]{LeeYang.1956}
Lee, T.~D., \& Yang, C.~N. 1956, Physical Review, 104, 254

\bibitem[{Lindhard \& Scharff(1961)}]{Lindhard.1961}
Lindhard, J., \& Scharff, M. 1961, Physical Review, 124, 128

\bibitem[{Mathis {et~al.}(1983)Mathis, Mezger, \& Panagia}]{Mathis.1983}
Mathis, J.~S., Mezger, P.~G., \& Panagia, N. 1983, \aa, 128, 212

\bibitem[{McGuire {et~al.}(2016)McGuire, Carroll, Loomis, Finneran, Jewell,
  Remijan, \& Blake}]{McGuire.2016}
McGuire, B.~A., Carroll, P.~B., Loomis, R.~A., {et~al.} 2016, Science, 352,
  1449

\bibitem[{Meinert {et~al.}(2016)Meinert, Myrgorodska, Marcellus, Buhse, Nahon,
  Hoffmann, d’Hendecourt, \& Meierhenrich}]{Meinert.2016}
Meinert, C., Myrgorodska, I., Marcellus, P.~d., {et~al.} 2016, Science, 352,
  208

\bibitem[{{Meng} {et~al.}(2019){Meng}, {S{\'a}nchez-Monge}, {Schilke},
  {Padovani}, {Marcowith}, {Ginsburg}, {Schmiedeke}, {Schw{\"o}rer}, {DePree},
  {Veena}, \& {M{\"o}ller}}]{meng2019}
{Meng}, F., {S{\'a}nchez-Monge}, {\'A}., {Schilke}, P., {et~al.} 2019, \aap,
  630, A73

\bibitem[{Miller \& Urey(1959)}]{MillerUrey.1959}
Miller, S.~L., \& Urey, H.~C. 1959, Science, 130, 245

\bibitem[{Miotello {et~al.}(2014)Miotello, Testi, Lodato, Ricci, Rosotti,
  Brooks, Maury, \& Natta}]{Miotello.2014}
Miotello, A., Testi, L., Lodato, G., {et~al.} 2014, \aa, 567, A32

\bibitem[{Modica {et~al.}(2018)Modica, Martins, Meinert, Zanda, \&
  d’Hendecourt}]{Modica.2018}
Modica, P., Martins, Z., Meinert, C., Zanda, B., \& d’Hendecourt, L. L.~S.
  2018, The Astrophysical Journal, 865

\bibitem[{Modica {et~al.}(2014)Modica, Meinert, Marcellus, Nahon, Meierhenrich,
  \& d'Hendecourt}]{Modica.2014}
Modica, P., Meinert, C., Marcellus, P.~d., {et~al.} 2014, The Astrophysical
  Journal, 788, 79

\bibitem[{{Morlino} {et~al.}(2021){Morlino}, {Blasi}, {Peretti}, \&
  {Cristofari}}]{morlino2021}
{Morlino}, G., {Blasi}, P., {Peretti}, E., \& {Cristofari}, P. 2021, \mnras,
  504, 6096

\bibitem[{Naaman {et~al.}(2018)Naaman, Paltiel, \& Waldeck}]{Naaman.2018}
Naaman, R., Paltiel, Y., \& Waldeck, D.~H. 2018, CHIMIA International Journal
  for Chemistry, 72, 394

\bibitem[{Naaman \& Waldeck(2012)}]{Naaman.2012}
Naaman, R., \& Waldeck, D.~H. 2012, The Journal of Physical Chemistry Letters,
  3, 2178

\bibitem[{{Neufeld} \& {Wolfire}(2017)}]{neufeld2017}
{Neufeld}, D.~A., \& {Wolfire}, M.~G. 2017, \apj, 845, 163

\bibitem[{Oba {et~al.}(2023)Oba, Takano, Dworkin, \& Naraoka}]{Oba.2023}
Oba, Y., Takano, Y., Dworkin, J.~P., \& Naraoka, H. 2023, Nature
  Communications, 14, 3107

\bibitem[{Oba {et~al.}(2016)Oba, Takano, Watanabe, \& Kouchi}]{Oba.2016}
Oba, Y., Takano, Y., Watanabe, N., \& Kouchi, A. 2016, The Astrophysical
  Journal Letters, 827, L18

\bibitem[{{Ozturk} {et~al.}(2023{\natexlab{a}}){Ozturk}, {Bhowmick}, {Kapon},
  {Sang}, {Kumar}, {Paltiel}, {Naaman}, \& {Sasselov}}]{Ozturk.2023a}
{Ozturk}, S.~F., {Bhowmick}, D.~K., {Kapon}, Y., {et~al.} 2023{\natexlab{a}},
  Nature Communications, 14, 6351

\bibitem[{{Ozturk} {et~al.}(2023{\natexlab{b}}){Ozturk}, {Liu}, {Sutherland},
  \& {Sasselov}}]{Ozturk.2023b}
{Ozturk}, S.~F., {Liu}, Z., {Sutherland}, J.~D., \& {Sasselov}, D.~D.
  2023{\natexlab{b}}, Science Advances, 9, eadg8274

\bibitem[{{Ozturk} \& {Sasselov}(2022)}]{Ozturk.2022}
{Ozturk}, S.~F., \& {Sasselov}, D.~D. 2022, Proceedings of the National Academy
  of Science, 119, e2204765119

\bibitem[{Padovani {et~al.}(2009)Padovani, Galli, \& Glassgold}]{Padovani.2009}
Padovani, M., Galli, D., \& Glassgold, A.~E. 2009, Astronomy \& Astrophysics,
  501, 619

\bibitem[{Padovani {et~al.}(2024)Padovani, Galli, Scarlett, Grassi, Rehill,
  Zammit, Bray, \& Fursa}]{Padovani.2024}
Padovani, M., Galli, D., Scarlett, L.~H., {et~al.} 2024, Astronomy \&
  Astrophysics, 682, A131

\bibitem[{{Padovani} {et~al.}(2015){Padovani}, {Hennebelle}, {Marcowith}, \&
  {Ferri{\`e}re}}]{padovani2015}
{Padovani}, M., {Hennebelle}, P., {Marcowith}, A., \& {Ferri{\`e}re}, K. 2015,
  \aap, 582, L13

\bibitem[{{Padovani} {et~al.}(2016){Padovani}, {Marcowith}, {Hennebelle}, \&
  {Ferri{\`e}re}}]{padovani2016}
{Padovani}, M., {Marcowith}, A., {Hennebelle}, P., \& {Ferri{\`e}re}, K. 2016,
  \aap, 590, A8

\bibitem[{{Padovani} {et~al.}(2019){Padovani}, {Marcowith},
  {S{\'a}nchez-Monge}, {Meng}, \& {Schilke}}]{padovani2019}
{Padovani}, M., {Marcowith}, A., {S{\'a}nchez-Monge}, {\'A}., {Meng}, F., \&
  {Schilke}, P. 2019, \aap, 630, A72

\bibitem[{{Parizot} {et~al.}(2004){Parizot}, {Marcowith}, {van der Swaluw},
  {Bykov}, \& {Tatischeff}}]{parizot2004}
{Parizot}, E., {Marcowith}, A., {van der Swaluw}, E., {Bykov}, A.~M., \&
  {Tatischeff}, V. 2004, \aap, 424, 747

\bibitem[{Parker {et~al.}(2023)Parker, McLain, Glavin, Dworkin, Elsila, Aponte,
  Naraoka, Takano, Tachibana, Yabuta, Yurimoto, Sakamoto, Yada, Nishimura,
  Nakato, Miyazaki, Yogata, Abe, Okada, Usui, Yoshikawa, Saiki, Tanaka,
  Nakazawa, Tsuda, Terui, Noguchi, Okazaki, Watanabe, \&
  Nakamura}]{Parker.2023}
Parker, E.~T., McLain, H.~L., Glavin, D.~P., {et~al.} 2023, Geochimica et
  Cosmochimica Acta, 347, 42

\bibitem[{Patty {et~al.}(2021)Patty, Kühn, Lambrev, Spadaccia, Hoeijmakers,
  Keller, Mulder, Pallichadath, Poch, Snik, Stam, Pommerol, \&
  Demory}]{Patty.2021}
Patty, C. H.~L., Kühn, J.~G., Lambrev, P.~H., {et~al.} 2021, Astronomy \&
  Astrophysics, 651, A68

\bibitem[{Pfandzelter {et~al.}(2003)Pfandzelter, Winter, Urazgil’din, \&
  Rösler}]{Pfandzelter.2003}
Pfandzelter, R., Winter, H., Urazgil’din, I., \& Rösler, M. 2003, Physical
  Review B, 68, 165415

\bibitem[{{Phan} {et~al.}(2020){Phan}, {Gabici}, {Morlino}, {Terrier}, {Vink},
  {Krause}, \& {Menu}}]{phan2020}
{Phan}, V.~H.~M., {Gabici}, S., {Morlino}, G., {et~al.} 2020, \aap, 635, A40

\bibitem[{{Phan} {et~al.}(2021){Phan}, {Schulze}, {Mertsch}, {Recchia}, \&
  {Gabici}}]{phan2021}
{Phan}, V. H.~M., {Schulze}, F., {Mertsch}, P., {Recchia}, S., \& {Gabici}, S.
  2021, \prl, 127, 141101

\bibitem[{Pizzarello \& Cronin(1998)}]{Pizzarello.1998}
Pizzarello, S., \& Cronin, J.~R. 1998, Nature, 394, 236

\bibitem[{Pizzarello \& Huang(2005)}]{Pizzarello.2005}
Pizzarello, S., \& Huang, Y. 2005, Geochimica et Cosmochimica Acta, 69, 599

\bibitem[{Potiszil {et~al.}(2023)Potiszil, Ota, Yamanaka, Sakaguchi, Kobayashi,
  Tanaka, Kunihiro, Kitagawa, Abe, Miyazaki, Nakato, Nakazawa, Nishimura,
  Okada, Saiki, Tanaka, Terui, Tsuda, Usui, Watanabe, Yada, Yogata, Yoshikawa,
  \& Nakamura}]{Potiszil.2023}
Potiszil, C., Ota, T., Yamanaka, M., {et~al.} 2023, Nature Communications, 14,
  1482

\bibitem[{Purcell(1979)}]{Purcell.1979}
Purcell, E.~M. 1979, \apj, 231, 404

\bibitem[{Ray {et~al.}(1999)Ray, Ananthavel, Waldeck, \& Naaman}]{Ray.1999}
Ray, K., Ananthavel, S.~P., Waldeck, D.~H., \& Naaman, R. 1999, Science, 283,
  814

\bibitem[{{Redaelli} {et~al.}(2021){Redaelli}, {Sipil{\"a}}, {Padovani},
  {Caselli}, {Galli}, \& {Ivlev}}]{redaelli2021}
{Redaelli}, E., {Sipil{\"a}}, O., {Padovani}, M., {et~al.} 2021, \aap, 656,
  A109

\bibitem[{Rivilla {et~al.}(2022)Rivilla, Jiménez-Serra, Martín-Pintado,
  Colzi, Tercero, Vicente, Zeng, Martín, Concepción, Bizzocchi, Melosso,
  Rico-Villas, \& Requena-Torres}]{Rivilla.2022}
Rivilla, V.~M., Jiménez-Serra, I., Martín-Pintado, J., {et~al.} 2022,
  Frontiers in Astronomy and Space Sciences, 9, 876870

\bibitem[{Rivilla {et~al.}(2023)Rivilla, Sanz-Novo, Jiménez-Serra,
  Martín-Pintado, Colzi, Zeng, Megías, López-Gallifa, Martínez-Henares,
  Massalkhi, Tercero, Vicente, Martín, Andrés, Requena-Torres, \&
  Alonso}]{Rivilla.2023}
Rivilla, V.~M., Sanz-Novo, M., Jiménez-Serra, I., {et~al.} 2023, The
  Astrophysical Journal Letters, 953, L20

\bibitem[{Rosenberg(2019)}]{Rosenberg.2019}
Rosenberg, R.~A. 2019, Symmetry, 11, 528

\bibitem[{Rosenberg {et~al.}(2008)Rosenberg, Haija, \& Ryan}]{Rosenberg.2008}
Rosenberg, R.~A., Haija, M.~A., \& Ryan, P.~J. 2008, Physical Review Letters,
  101, 178301

\bibitem[{Rosenberg {et~al.}(2015)Rosenberg, Mishra, \&
  Naaman}]{Rosenberg.2015}
Rosenberg, R.~A., Mishra, D., \& Naaman, R. 2015, Angewandte Chemie
  International Edition, 54, 7295

\bibitem[{Rudd {et~al.}(1983)Rudd, DuBois, Toburen, Ratcliffe, \&
  Goffe}]{Rudd.1983}
Rudd, M.~E., DuBois, R.~D., Toburen, L.~H., Ratcliffe, C.~A., \& Goffe, T.~V.
  1983, Physical Review A, 28, 3244

\bibitem[{{Sabatini} {et~al.}(2023){Sabatini}, {Bovino}, \&
  {Redaelli}}]{sabatini2023}
{Sabatini}, G., {Bovino}, S., \& {Redaelli}, E. 2023, \apjl, 947, L18

\bibitem[{Savage \& Bohlin(1979)}]{Savage.1979}
Savage, B.~D., \& Bohlin, R.~C. 1979, The Astrophysical Journal, 229, 136

\bibitem[{{Scherer} {et~al.}(2008){Scherer}, {Fichtner}, {Ferreira},
  {B{\"u}sching}, \& {Potgieter}}]{scherer2008}
{Scherer}, K., {Fichtner}, H., {Ferreira}, S.~E.~S., {B{\"u}sching}, I., \&
  {Potgieter}, M.~S. 2008, \apjl, 680, L105

\bibitem[{Shen {et~al.}(2008)Shen, Draine, \& Johnson}]{Shen.2008ApJ}
Shen, Y., Draine, B.~T., \& Johnson, E.~T. 2008, \apj, 689, 260

\bibitem[{Soai {et~al.}(1995)Soai, Kawasaki, \& Matsumoto}]{Soai.1995}
Soai, K., Kawasaki, T., \& Matsumoto, A. 1995, Nature, 14, 70

\bibitem[{Sorrell(1994)}]{Sorrell.1994}
Sorrell, W.~H. 1994, Monthly Notices of the Royal Astronomical Society, 268, 40

\bibitem[{{Sparks} {et~al.}(2015){Sparks}, {Hough}, \&
  {Kolokolova}}]{Sparks.2015}
{Sparks}, W., {Hough}, J., \& {Kolokolova}, L. 2015, in Polarimetry of Stars
  and Planetary Systems, ed. L.~{Kolokolova}, J.~{Hough}, \& A.-C.
  {Levasseur-Regourd} ((New York: Cambridge Univ. Press)), 462

\bibitem[{Sparks {et~al.}(2009{\natexlab{a}})Sparks, Hough, Kolokolova, Germer,
  Chen, DasSarma, DasSarma, Robb, Manset, Reid, Macchetto, \&
  Martin}]{Sparks.2009md9}
Sparks, W., Hough, J., Kolokolova, L., {et~al.} 2009{\natexlab{a}}, Journal of
  Quantitative Spectroscopy and Radiative Transfer, 110, 1771

\bibitem[{Sparks {et~al.}(2009{\natexlab{b}})Sparks, Hough, Germer, Chen,
  DasSarma, DasSarma, Robb, Manset, Kolokolova, Reid, Macchetto, \&
  Martin}]{Sparks.2009}
Sparks, W.~B., Hough, J., Germer, T.~A., {et~al.} 2009{\natexlab{b}},
  Proceedings of the National Academy of Sciences, 106, 7816

\bibitem[{Sternheimer {et~al.}(1984)Sternheimer, Berger, \&
  Seltzer}]{1984ADNDT..30..261S}
Sternheimer, R.~M., Berger, M.~J., \& Seltzer, S.~M. 1984, Atomic Data and
  Nuclear Data Tables, 30, 261

\bibitem[{Sullivan {et~al.}(2016)Sullivan, Boamah, Shulenberger, Chapman,
  Atkinson, Boyer, \& Arumainayagam}]{Sullivan.2016}
Sullivan, K.~K., Boamah, M.~D., Shulenberger, K.~E., {et~al.} 2016, Monthly
  Notices of the Royal Astronomical Society, 460, 664

\bibitem[{Takayanagi(1973)}]{Takayanagi.1973}
Takayanagi, K. 1973, PASJ

\bibitem[{Tassinari {et~al.}(2019)Tassinari, Steidel, Paltiel, Fontanesi,
  Lahav, Paltiel, \& Naaman}]{Tassinari.2019}
Tassinari, F., Steidel, J., Paltiel, S., {et~al.} 2019, Chemical Science, 10,
  5246

\bibitem[{Unguris {et~al.}(1982)Unguris, Pierce, Galejs, \&
  Celotta}]{Unguris.1982}
Unguris, J., Pierce, D.~T., Galejs, A., \& Celotta, R.~J. 1982, Physical Review
  Letters, 49, 72

\bibitem[{Vaillancourt {et~al.}(2020)Vaillancourt, Andersson, Clemens, Piirola,
  Hoang, Becklin, \& Caputo}]{Vaillancourt:2020ch}
Vaillancourt, J.~E., Andersson, B.-G., Clemens, D.~P., {et~al.} 2020, \apj,
  905, 0

\bibitem[{{Vaupr{\'e}} {et~al.}(2014){Vaupr{\'e}}, {Hily-Blant}, {Ceccarelli},
  {Dubus}, {Gabici}, \& {Montmerle}}]{vaupre2014}
{Vaupr{\'e}}, S., {Hily-Blant}, P., {Ceccarelli}, C., {et~al.} 2014, \aap, 568,
  A50

\bibitem[{{Vieu} {et~al.}(2022){Vieu}, {Gabici}, {Tatischeff}, \&
  {Ravikularaman}}]{vieu2022}
{Vieu}, T., {Gabici}, S., {Tatischeff}, V., \& {Ravikularaman}, S. 2022,
  \mnras, 512, 1275

\bibitem[{Weingartner \& Draine(2001)}]{WeingartnerDraine.2001a}
Weingartner, J.~C., \& Draine, B.~T. 2001, \apjs, 134, 263

\bibitem[{Whittet {et~al.}(2008)Whittet, Hough, Lazarian, \&
  Hoang}]{Whittet.2008}
Whittet, D. C.~B., Hough, J.~H., Lazarian, A., \& Hoang, T. 2008, \apj, 674,
  304

\bibitem[{Wu {et~al.}(1957)Wu, Ambler, Hayward, Hoppes, \& Hudson}]{Wu.1957}
Wu, C.~S., Ambler, E., Hayward, R.~W., Hoppes, D.~D., \& Hudson, R.~P. 1957,
  Physical Review, 105, 1413

\bibitem[{Wu {et~al.}(2024)Wu, Anderson, Watkins, Arora, Barnes, Padovani,
  Shingledecker, Arumainayagam, \& Battat}]{Wu.2024}
Wu, Q.~T., Anderson, H., Watkins, A.~K., {et~al.} 2024, ACS Earth and Space
  Chemistry, 8, 79

\bibitem[{Zhukovska {et~al.}(2008)Zhukovska, Gail, \&
  Trieloff}]{Zhukovska:2008p3096}
Zhukovska, S., Gail, H.-P., \& Trieloff, M. 2008, \aa, 479, 453

\bibitem[{Ziegler(1999)}]{1999JAP....85.1249Z}
Ziegler, J.~F. 1999, JAP, 85, 1249

\end{thebibliography}

\appendix
\section{$H_{2}$ Ionization Cross Section by Cosmic Rays}\label{apdx:cross-section}
The $H_{2}$ ionization cross-section by CR electrons is given by \citep{Padovani.2009}:
\bea
\sigma_{\H_{2}}^{\rm CRe}=4\pi a_{0}^{2}ZF(t)G(t)\left(\frac{I(\H)}{I(\H_{2})}\right)^{2},\label{eq:sigma_ion}
\ena
where $a_{0}=\hbar^{2}/(m_{e}e^{2})$ is the Bohr radius, $I(\H)=13.598$ eV is the hydrogen ionization potential, $t=E/I(\H_{2})$ with $Z=2$ for two electrons of a H$_{2}$ molecule, 
\bea
F(t)=\frac{1-t^{1-n}}{n-1}-\left(\frac{2}{1+t} \right)^{n/2}\frac{1-t^{1-n/2}}{n-2},\\
G(t)= \frac{1}{t}\left(A_{1}\ln t + A_{2} +\frac{A_{3}}{t}\right),\label{eq:FG}
\ena
where $n=2.4\pm 0.2, A_{1}=0.74\pm 0.02, A_{2}=0.87\pm 0.06, A_{3}=-0.60\pm 0.05$. 

The $\H_{2}$ ionization cross-section by CR protons is given by
\bea
\sigma_{\rm CRp}= \left(\sigma_{1}^{-1} + \sigma_{h}^{-1}\right)^{-1},
\ena
where
\bea
\sigma_{1} = 4\pi a_{0}^{2}Cx^{D},\\
\sigma_{h} = 4\pi a_{0}^{2}(A\ln(1+x)+ B)x^{-1},\
\ena
where $x = m_{e}E/(m_{\H}I(\H))$, and $A=0.71, B=1.63, C=0.51, D=1.24$ \citep{Rudd.1983}.

The cross-section for the electron capture of CR protons via the $p + H_{2}\rightarrow H + H^{+}_{2}$ is
\bea
\sigma_{\rm CRp}^{e.c}= 4\pi a_{0}^{2}AZ\left(\frac{I(\H)}{I(\H_{2})}\right)^{2}\frac{x^{2}}{(C + x^{B}+ Dx^{F})},\label{eq:sigma_CRp_ec}
\ena
where $m_{e}E/(m_{\H}I(\H))$, $A=1.044, B=2.88, C=0.016, D=0.136, F=5.86$ \citep{Rudd.1983}.

\section{Energy Loss of CR electrons and protons}\label{apdx:energy_loss}
\subsection{CR protons}
Consider the interaction of an incident CR ion (projectile) of charge $Z_{P}e$ and kinetic energy $E$ (velocity $\beta c$) with an electron of the target atom. Let $Z_{T,i}$ be the atomic number of target element $i$. The total energy loss of the incident ion per pathlength, which is usually referred to as {\it stopping power} of the material, is given by
\bea
\frac{dE_{\rm ion}}{dx}=\sum_{i} n_{i}S_{i},\label{eq:dEdx_ion}
\ena 
where $n_{i}$ is the atomic number density of element $i$ in the gas or dust, $S_{i}$ is the electronic stopping cross-section of element $i$ in units of $\eV \cm^{2}$, and the sum is taken over all elements $i$ present in the grain. For the $\H_{2}$ gas, $n_{i}=n_{H2}$. For the silicate grains of structure MgFeSiO$_{4}$, the summation is over all these elements with their respective fraction.

The Bethe-Bloch theory yields the following (see \citealt{1963ARNPS..13....1F}):
\bea
S_{i} =\frac{4\pi(Z_{P}e^{2})^{2}}{m_{e}c^{2}\beta^{2}}Z_{T,i} \left[\frac{1}{2}\ln \frac{2\gamma^{2}m_{e}c^{2}\beta^{2}T_{\max}}{I_{i}^{2}}-\beta^{2}-\frac{C}{Z_{T,i}}-\frac{\delta}{2}\right],\label{eq:Sion}~~~
\ena
where $T_{\max}$ is the maximum energy transferred from the ion to the atomic electron, and $I_{i}$ is the mean excitation energy of element $i$. Here $C/Z_{T,i}$ is the shell correction term, and $\delta/2$ is the density correction term.

In binary collisions, $T_{\max}$ is given by
\bea
T_{\max}=\frac{2\gamma^{2}M_{P}^{2}m_{e}c^{2}\beta^{2}}{m_{e}^{2}+M_{P}^{2}+2\gamma m_{e}M_{P}},\label{eq:Tmax}
\ena
where $M_{P}$ is the atomic mass of the projectile.

In the case $2\gamma m_{e}/M_{P} \ll 1$ (heavy ion or not very high $\gamma$) we have $T_{\max}=2\gamma^{2}\beta^{2}m_{e}c^{2}$. Thus Equation (\ref{eq:dE-dx}) can be rewritten as
\bea
S_{i}=\frac{4\pi(Z_{P}e^{2})^{2}}{m_{e}c^{2}\beta^{2}}Z_{T,i}\left[\ln \frac{2m_{e}c^{2}\beta^{2}}{I_{i}}+\mathcal{L}(\beta)\right],\label{eq:dEidx}
\ena
where
\bea
\mathcal{L}(\beta)=\ln \frac{1}{1-\beta^{2}}-\beta^{2}-\frac{C}{Z_{T,i}}-\frac{\delta}{2}.\label{eq:Lbeta}
\ena
where the shell correction term $C/Z_{T,i}$ is maximum at $\sim 0.3$ and becomes negligible for $E >10$ MeV/amu. The density correction term $\delta/2$ is important and increases to above $1$ for $E> 5$ GeV/amu  (see \citealt{1999JAP....85.1249Z} for more details). Here we calculate $\delta/2$ using an analytical fit from \cite{1984ADNDT..30..261S}.

For energy below 1 MeV, the energy loss is given by (\citealt{Lindhard.1961})
\bea
S_{i}= 8\pi e^{2}a_{0}Z_{P}^{1/6}\frac{Z_{P}Z_{T}}{Z}\frac{\beta c}{v_{Bohr}}
\ena
where 
\bea
Z	 = \left(Z_{P}^{2/3} + Z_{T}^{2/3}\right)^{3/2}
\ena
and $v_{\rm Bohr}= e^{2}/\hbar$.

\subsection{CR electrons}
The energy loss of a CR electron with kinetic energy $E$ per pathlength in the dust grain due to electronic excitations is equal to (see \citealt{Hoang.2015})
\bea
\frac{dE_{{\rm el}}}{dx}=\sum_{i} n_{i}S_{{\rm el},i},\label{eq:dEdx_el}\\
\ena
where $S_{{\rm el},i}$ is the energy loss (or stopping power) caused by atom $i$ of charge number $Z_{T,i}$ given by 
\bea
S_{{\rm el},i}=\frac{4\pi e^{4}}{m_{e}c^{2}\beta^{2}}Z_{T,i}\left[\frac{1}{2}\ln \left(\frac{E^{2}}{I_{i}^{2}}\frac{\gamma+1}{2}\right)+\frac{\mathcal{F}(\gamma)}{2}\right],\label{eq:Sel}
\ena
where 
\bea
\mathcal{F}(\gamma)=-\left(\frac{2}{\gamma}-\frac{1}{\gamma^{2}}\right)\ln 2 +\frac{1}{\gamma^{2}}+\frac{1}{8}\left(1-\frac{1}{\gamma}\right)^{2}.\label{eq:Fgamma}
\ena

\section{Continous Slowing Down Approximation (CSDA) Model for CR Transport in Protostellar Cores}\label{apdx:CRtransport}
When penetrating the protostellar core, CRs loose energy due to interaction with $\H_{2}$. The CR energy loss is described by the loss function 
\bea
L(E)=\frac{1}{n_{\H_{2}}}\frac{dE}{dx},\label{eq:LE}
\ena
where $n_{\H_{2}}$ is the $\H_{2}$ density of the gas.

Following the CSDA model of CR transport \citep{Takayanagi.1973}, the flux of CR particles is conserved such that
\bea
j_{k}(E)dE = j_{k}(E_{0})dE_{0}.\label{eq:jdE_jdE0}
\ena

The decrease in the energy of CRs after a pathlenght $dx$ is given by
\bea
dE = L_{k}(E) n_{\H_{2}}dx=L_{k}(E)dN,
\ena
where $dN = n_{\H_{2}}dx$, which implies
\bea
\frac{dE_{0}}{dE}=\frac{L_{k}(E_{0})}{L_{k}(E)}.
\ena 

From the above equations, one can obtain
\bea
j_{k}(E) = j_{k}(E_{0})\frac{L_{k}(E_{0})}{L_{k}(E)},\label{eq:jE_jE0}
\ena
which links the original flux and the final flux.

The total column density required to decrease from $E_{0}$ to $E$ is given by
\bea
N_{\H_{2}} = \int_{0}^{N_{\H_{2}}}dN = \int_{E_{0}}^{E}\frac{dE'}{L_{k}(E')}=n_{H_{2}}\left[R(E_{0})-R(E)\right],\label{eq:NH2_slowing}
\ena
where the range $dR=\frac{dE'}{n_{\H_{2}}L_{k}(E')}$.

For a given $N_{\H_{2}}$, the above equation links the original energy and the final energy $E$. To get the exact relationship, we calculate $N_{\H_{2}}$ for a grid of $E_{0}$ and $E$. Then, we select a contour level $N_{\H_{2}}$ and infer the parameters $E,E_{0}$ which is then fitted by an analytical function
\bea
E_{0} =\left(cE^{b} + \frac{N_{\H_{2}}}{N_{0}} \right)^{1/b},\label{eq:E0_E} 
\ena
where $b,c,N_{0}$ are the fitting parameters.

To calculate the CR spectrum at a specific column density, $j_{k}(E,N_{\H_{2}})$, we use Eq. \ref{eq:jE_jE0} where $E_{0}$ is given by Eq.\ref{eq:E0_E}.
\begin{figure}
	\includegraphics[width=0.9\textwidth]{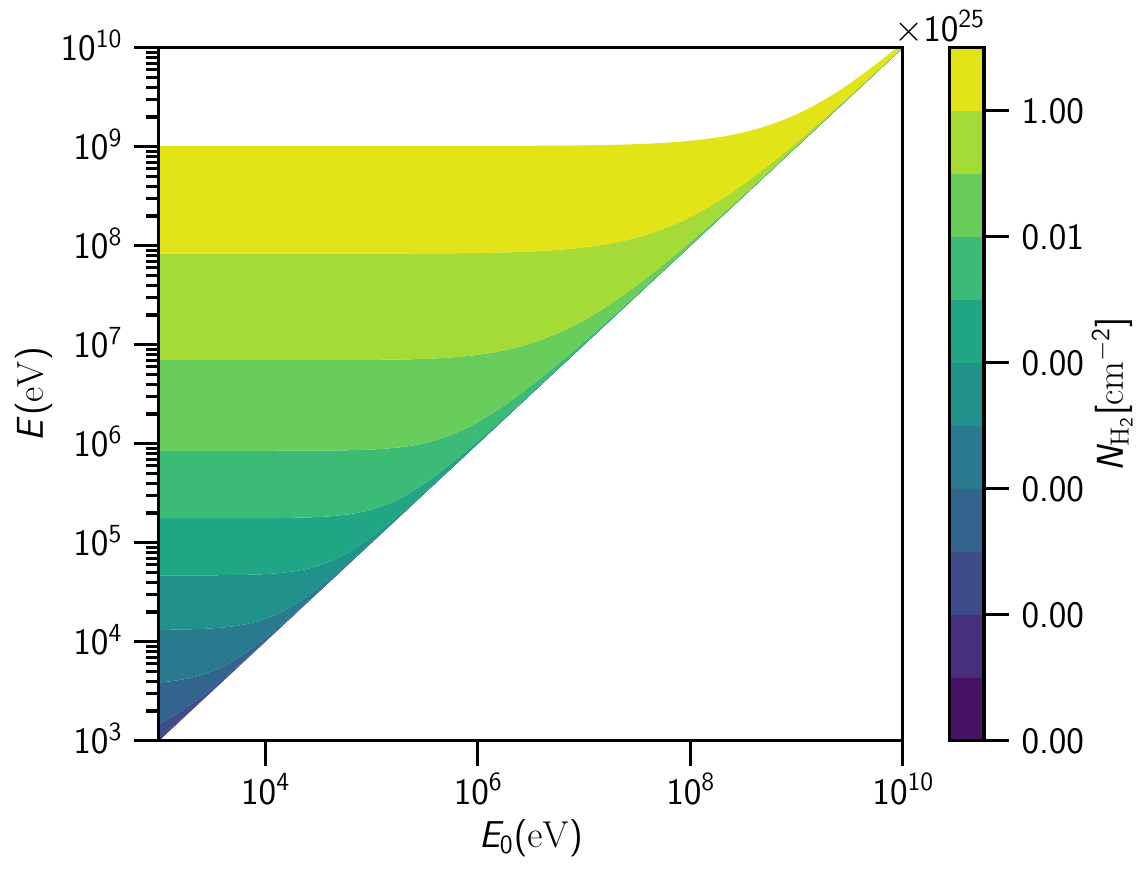}
	\caption{Contours of $N_{\H_{2}}$ as functions of $E_{0}$ and $E$ for energy loss of CR electrons in the $\H_{2}$ gas. For a fixed contour level $N_{\H_{2}}$, there is a unique relationship between the energy of CRs after passing through that column density, $E$, to the initial energy of CRs $E_{0}$.}
	\label{fig:contour_NH2}
\end{figure}

Figure \ref{fig:contour_NH2} shows the contour map of $N_{\H_{2}}$ for CR electrons. For each contour of a given $N_{\H_{2}}$, one can obtain a relationship between the original energy $E_{0}$ to the energy $E$ of the CR particle that already traverses the column $N_{\H_{2}}$.

\end{document}